\documentclass[acmsmall]{acmart}

\usepackage{alltt}
\usepackage{epsfig}
\usepackage{verbatim}
\usepackage{fancybox}
\usepackage{multirow}
\usepackage{colortbl}
\usepackage{rotating}
\usepackage{algorithm}
\usepackage{algorithmic}

\usepackage{color}
\usepackage[ampersand]{easylist}
\usepackage{cancel}
\usepackage{amsmath}
\usepackage{pgfplots}
\pgfplotsset{compat=newest}
\usepackage{tikz}
\usepackage{xcolor}
\usepackage[textsize=scriptsize]{todonotes}

\usepackage{gensymb}
\definecolor{LightCyan}{rgb}{0.88,1,1}

\AtBeginDocument{%
  \providecommand\BibTeX{{%
    \normalfont B\kern-0.5em{\scshape i\kern-0.25em b}\kern-0.8em\TeX}}}

\setcopyright{acmcopyright}
\acmJournal{TSAS}
\acmYear{2022} \acmVolume{1} \acmNumber{1} \acmArticle{1} \acmMonth{1} \acmPrice{15.00}\acmDOI{10.1145/3516524}

\begin{document}

\newcommand{\etal}{{\em et al. }}
\newcommand{\virus}{{\em COVID-19\ }}
\newcommand{\smu}{{\em SMU\ }}
\newcommand{\nus}{{\em NUS\ }}
\newcommand{\ndorm}{{\em NUS\_Dorm\ }}
\newcommand{\amherst}{{\em UMASS\ }}
\newcommand{\adorm}{{\em UMASS\_Dorm\ }}

\newcommand{\etals}{{\em et al.}}
\newcommand{\viruss}{{\em COVID-19}}
\newcommand{\smus}{{\em SMU}}
\newcommand{\nuss}{{\em NUS}}
\newcommand{\amhersts}{{\em UMASS}}
\newcommand{\adorms}{{\em UMASS\_Dorm}}
\newcommand{\ndorms}{{\em NUS\_Dorm}}

\title{Analyzing the Impact of COVID-19 Control Policies on Campus Occupancy and Mobility via WiFi Sensing}

\author{Camellia Zakaria}
\affiliation{%
  \institution{University of Massachusetts Amherst}
  \country{USA}}
\email{nurcamellia@cs.umass.edu}

\author{Amee Trivedi}
\affiliation{%
  \institution{University of Massachusetts Amherst}
  \country{USA}}
\email{amee@cs.umass.edu}

\author{Emmanuel Cecchet}
\affiliation{%
  \institution{University of Massachusetts Amherst}
  \country{USA}}
\email{cecchet@cs.umass.edu}

\author{Michael Chee}
\affiliation{%
  \institution{Duke-NUS Medical School}
  \country{Singapore}}
\email{michael.chee@nus.edu.sg}

\author{Prashant Shenoy}
\affiliation{%
  \institution{University of Massachusetts Amherst}
  \country{USA}}
\email{shenoy@cs.umass.edu}

\author{Rajesh Balan}
\affiliation{%
  \institution{Singapore Management University}
  \country{Singapore}}
\email{rajesh@smu.edu.sg}

\renewcommand{\shortauthors}{Zakaria, Trivedi, Cecchet, Chee, Shenoy and Balan, et al.}
\begin{abstract}
Mobile sensing has played a key role in providing digital solutions to aid with \virus containment policies, primarily to automate contact tracing and social distancing measures. As more and more countries reopen from lockdowns, there remains a pressing need to minimize crowd movements and interactions, particularly in enclosed spaces. Many \virus technology solutions leverage positioning systems, generally using Bluetooth and GPS, and can theoretically be adapted to monitor safety compliance within dedicated environments. However, they may not be the ideal modalities for indoor positioning. This paper conjectures that analyzing user occupancy and mobility via deployed WiFi infrastructure can help institutions monitor and maintain safety compliance according to the public health guidelines. Using smartphones as a proxy for user location, our analysis demonstrates how coarse-grained WiFi data can sufficiently reflect the indoor occupancy spectrum when different \virus policies were enacted. Our work analyzes staff and students' mobility data from three university campuses. Two of these campuses are in Singapore, and the third is in the Northeastern United States. Our results show that online learning, split-team, and other space management policies effectively lower occupancy. However, they do not change the mobility for individuals transitioning between spaces. We demonstrate how this data source can be a practical application for institutional crowd control and discuss the implications of our findings for policy-making.
\end{abstract}



\maketitle

\section{Introduction}
Mobile sensing is increasingly employed to provide community support in security and safety \cite{guo2015mobile,ma2014opportunities} and, very quickly, proven helpful in the recent \virus global pandemic. For example, countries such as Singapore and South Korea promptly adopted mobile apps and sensors for various pandemic responses, including contact tracing and guiding social distancing policies \cite{tracetogether,safeentry,park2020coronavirus}. As our understanding of \virus constantly expands, data-driven analytics have paved the way to guide safe management measures by governments worldwide \cite{edb2020,osha2020}. Overall, these efforts have aided in revising health guidelines and even re-openings to crowds \cite{dar2020applicability}. Unfortunately, monitoring the safety compliance of these policies can be a struggle for agencies, organizations, and institutions as they rapidly change with new findings.

Sensing location information is fundamental in digital solutions responding to \viruss. For example, users' current location within communal spaces infers crowd density, allowing operations management to mobilize social distancing measures. Their location histories help identify places at risk of virus exposure in contact tracing procedures. Much research analyzing location data to combat \virus has been conducted over this short term. Badr \etal analyzed mobile phone signals from cell towers in 25 U.S. counties and found that mobility patterns strongly correlate with the virus spread \cite{badr2020association}. Positioning systems using Bluetooth and GPS immediately became pivotal in digital contact-tracing apps to stop the viral spread \cite{dar2020applicability}. Besides these mobile apps, crowd monitoring solutions (CMS) are also relevant to monitor gathering in enclosed spaces \cite{jain2019scaling,ayyalasomayajula2018bloc,xu2019adarf}. Recent efforts utilizing CMS for \virus include video images \cite{shorfuzzaman2021towards,barnawi2021artificial,kumar2021drone} and Bluetooth \cite{zhen2021ai}. Fundamentally, these solutions can procedurally track occupants' digital footprints of their physical whereabouts. Bluetooth, drones, or thermal imaging cameras are less privacy-invasive than typical video surveillance. However, institutions may lack operational readiness, requiring additional device deployment in dedicated environments. A feasible solution is leveraging the existing WiFi network typically available to occupants in educational institutions \cite{trivedi2021wifitrace,zhang2021wlan}. These works proposed inference methods to identify at-risk occupants or superspreader events based on their associations with specific access points (APs) and contact graphs to reveal potential directional interactions between individuals. The effectiveness of these techniques was demonstrated through simulated examples.

In the same vein, our work seeks to investigate if location data that is passively sensed from existing WiFi infrastructure can, in fact, show the real-world effects of various \virus policies on institutions when the pandemic broke out. Our goal is to show that coarse-grained WiFi data can sufficiently reflect the spectrum in crowd change when different control policies are implemented over time within an institution. Monitoring crowd change at three-level granularity (i.e., area, floor, and building) can contribute to the institutions’ ability to report on actual space utilization in addition to specific details, such as total building capacity, buildings' floor-by-floor, and area metrics. Besides being alerted of densely populated parts of the building, monitoring building occupancy and mobility is critical in informing the maximum range of occupancy and assessing exposure and vulnerabilities from occupants' mobility across shared open spaces. We present findings from analyzing anonymized and aggregated students and staff's location data through collecting WiFi logs directly from deployed campus WiFi infrastructures across three universities\footnote{Privacy and ethical considerations for our work are defined in Section \ref{sec:privacyEthicalConsiderations}.}. Two of these campuses are in Singapore, and one is in the northeastern part of the United States of America. We employ two key measures, 1) \textit{occupancy} -- the number of users represented by unique smartphone device counts in each building and 2) \textit{mobility} -- the order of buildings visited by the user.

This analysis comprises data during the initial phases of \virus for all universities in both countries. Our results show occupancy and mobility changing with different levels of restrictions. For example, we observed a reduction in occupancy when online learning was gradually introduced to some undergraduate classes. However, for on-campus staff and students, their mobility rate remained high. In contrast, a significant decrease in mobility rate is observed during the remote learning period. It is important to emphasize that this work neither solves a prediction problem nor proposes an alternative to digital contact tracing. Instead, our findings show the efficacy of WiFi sensing technique in regulating institutional compliance. The coarse-grained data analysis prevails as a data-driven approach to minimize congregation, moderate building occupancy, and crowd movement at an aggregated scale, preserving user privacy. Overall, this paper makes the following contributions:

\begin{enumerate}
    \item It provides one of the first detailed looks at the effect of \virus related policies on three campus populations across two different countries. Our results show how the \textit{occupancy} and \textit{mobility} metrics change across all three campuses as \virus quarantine policies, of increasing severity, are enacted.
    \item It provides evidence that passive monitoring of indoor occupancy and mobility patterns using WiFi data could be useful to administrators in deciding the appropriate safe distancing measures for similar disease outbreaks.
\end{enumerate}
\section{Background and Motivation}
In what follows, we summarize prior work relevant to digital \virus contact tracing and the practical use of WiFi sensing in the existing literature to monitor indoor occupancy and mobility for various applications.

\subsection{\virus College Outbreak} 
Close contact and congregations are discouraged during the pandemic since \virus can spread from person to person through respiratory droplets or by breathing in contaminated air. Some studies have reported that the virus can remain airborne over longer periods, especially in indoor spaces where it may likely have reduced ventilation \cite{cdcindoor}. The virus poses a direct threat to high population density areas, particularly work and education environments. Now, more than ever, precise indoor positioning capabilities are necessitated in these environments to model human mobility and identify areas at risk of disease spread.

One of the earliest virus outbreaks among college students was determined in March 2020, when 64 students were diagnosed positive with \virus as they returned to the United States from their Spring break \cite{lewis2020covid}. This outbreak was effectively controlled with swift cooperation and compliance between the university and public health officials. However, by July 2020, it was reported that 6,600 \virus cases linked to 270 colleges nationwide had begun to spread even before the Fall semester -- proving university campuses as highly vulnerable to the disease outbreak \cite{nyt2020}. As universities scrambled for campus closure, they were simultaneously pressured to respond to new public health strategies and requirements \cite{unesco2020education}, ranging from implementing restrictions on population movement to creating makeshift quarantine sites \cite{abc2020}. 

More recently, the New York Times reported on the success of major Singapore universities preventing the spread of \viruss. Specifically, Singapore universities developed monitoring systems to help residents comply with zone restrictions and other safety measures \cite{nytsg2021}. Simultaneously in the US, the Centers for Disease Control and Prevention (CDC) has provided several revisions on guidelines with which educational institutions must comply to reopen safely \cite{centers2020considerations}. Among these operation plans are decreasing occupancy in areas with limited indoor ventilation, staggering use, and restricting occupancy rate in shared spaces. The complexity of adapting rapidly changing rules and monitoring institutional compliance offers unprecedented mobile sensing opportunities to aid universities in making more informed decisions. 

\subsection{Digital Technologies for Monitoring \virus}
Since its outbreak, the public health response to \virus has quickly tapped into digital solutions to support various use-cases from public awareness to management protocols for different end-user groups such as the general public, facility administrators, and case investigators, respectively. Fundamentally, these efforts rely on localization and tracking methods to reduce the virus spread. This is particularly important as different countries, states, and even organizations enact a slew of policies they believe are best for them ~\cite{ILO}.

Several \virus trackers are available online, showing the number of infections per country \cite{worldometer,WHO_Stats,coronatracker,covid_JHU}, the spread rate \cite{surveillances2020epidemiological}, and possible ways to exit lockdown across the world \cite{yap2020covid}. At present, many crowd surveillance solutions used to support \virus safety compliance leverage video, footfall counter, and a combination of IoT sensors to measure crowd density, monitor crowd movement and the usage rate of facilities \cite{imda2020}. Real-time crowd density apps utilizing occupancy sensors, cameras, and ticket validations are also piloted in European cities to help commuters make well-informed decisions on the best travel routes and times \cite{eu2020}.

\subsubsection*{Smartphone-based Location Sensing}
One of the earliest analyses using mainly location data from the user's smartphone investigated the cluster of \virus cases in an office block in South Korea \cite{park2020coronavirus}. Confirmed case patients supplied their GPS information and were notified of the nearest screening center to get tested. Mobile phones quickly became key for exposure notifications and safety guidelines \cite{badr2020association,lee2020testing,dar2020applicability}. Simultaneously, Singapore mandated mobile check-ins at public spaces, for example, a shopping mall and every store within it, for residents to log their entries and exits using either a mobile phone or scanning their national identification card  \cite{safeentry}. While SafeEntry helps reduce manual logging effort for businesses, these procedures may be cumbersome to customers. Other solutions include digital contact-tracing mobile apps. Using Bluetooth and GPS, these apps track users' digital footprints of their whereabouts. Fundamentally, institutions can leverage such applications to manage their safety compliance. However, the sensing modalities pose several challenges. First, institutions may lack operational readiness, requiring the deployment of Bluetooth beacon devices in environments. Second, GPS struggles with indoor positioning from receiving inaccurate satellite signals. A viable solution is leveraging WiFi networks that are already deployed in institutions and are the most widely used facility by occupants.

\subsection{Leveraging WiFi-based Location Sensing}
\label{sec:whywifi}
Low user compliance in Bluetooth smartphone sensing for \virus has strongly motivated our research to understand the usefulness of WiFi sensing as a complementary modality for future contact tracing efforts.

\subsubsection{Indoor Crowd Monitoring Systems}
\label{sec:priorwifi}
The research community has long contributed to accurate and sustainable crowd monitoring systems (CMS). They include proposing the use of video surveillance \cite{jain2019scaling,papaioannou2016tracking}, RF \cite{ayyalasomayajula2018bloc} and Bluetooth devices \cite{xu2019adarf}. Recent CMS for \virus include using drones \cite{kumar2021drone}, cameras \cite{shorfuzzaman2021towards},  thermal imaging \cite{barnawi2021artificial} and Bluetooth beacons \cite{zhen2021ai}. Unfortunately, many challenges persist in achieving lasting real-world impact due to factors such as the cost of deployment and maintenance and cloud processing requirements.

WiFi sensing has lately revealed success in tracking indoor user mobility patterns for CMS \cite{duives2020enhancing}. Given widespread WiFi availability, these solutions are to understand consumer buying behaviors at shopping malls \cite{hwang2017process}, visitor analytics at airports and convention centers \cite{jayarajah2015need}, and students' health and work performance from deriving behavioral routines and social interactions on campus \cite{sen2014grumon,zakaria2019stressmon}. Recent applications for \virus include leveraging WiFi sensing to identify at-risk occupants or superspreader events based on their associations with specific access points (APs) \cite{trivedi2021wifitrace} and directional interactions between individuals \cite{zhang2021wlan}. The effectiveness of these techniques, however, was demonstrated through simulated examples.

Fundamentally, solutions to manage crowds measure the flow rate of people from one location to another; however, the location granularity for these modalities differ. For example, Bluetooth contact tracing mobile apps or camera-based CMS provide finer-grained inter-user distance trajectories. While WiFi sensing cannot provide such granularity (note: we discuss this as part of limitations in Section \ref{sec:conclusion}), picking up on nomadic movements between areas on a floor between buildings within a vicinity can produce digital footprints useful for identifying flagged paths of potential exposure within and across buildings. Through real-world examples, we show how simply monitoring indoor occupancy and mobility of occupants in three universities with WiFi can reflect the spectrum of crowd change when different \virus policies were introduced.

\subsubsection{Network-centric Sensing}
\label{sec:rationale}
Mobile phones have become ubiquitous in our daily lives, presenting many opportunities for behavioral sensing. Prior studies on using the ``phone as a sensor'' have gained new behavioral data \cite{palmer2013new,wang2014studentlife,harari2016using}. For example, smartphone use is prevalent among university students constituting the largest demographic of smartphone users \cite{crompton2018use}. At the same time, WiFi-based networks have achieved near-ubiquitous coverage in campus buildings and outdoor spaces, accommodating increasing student demands for learning and leisure activities \cite{wifi62020}. The ubiquitous availability of ``phone as a sensor'' for user behavior and WiFi-based networks as the sensing medium provides an ideal technology platform for our work. 

Mobile phone sensing comes in two forms: client-based and network-based. In client-based approaches, sensing is performed directly from the phone. Such sensing requires either OS support or a client-side mobile app (running continuously in the background) to perform sensing measurements. GPS is an example of a client-side sensing method using the OS (and GPS chip) to localize the phone. Bluetooth operates in similar ways and has been used for proximity sensing in digital contact tracing tools during the \virus pandemic. On the other
hand, network-based phone sensing involves using the wireless network for sensing measurements. Many of these approaches are passive -- which means they do not require active user involvement and can be performed via passive observations of the device. Network approaches have been used for indoor positioning systems (IPS) where multiple WiFi access points (APs) within range of the phone can ``triangulate'' the phone's location \cite{jayarajah2016livelabs}.

\begin{table} [!h]
\centering
\scalebox{0.8}{
\begin{tabular}{ |l|c|c| } \hline
\textbf{} & \textbf{Client-Centric} & \textbf{Network-Centric}\\
\textbf{Attributes} & \textbf{(Bluetooth/GPS)} & \textbf{(WiFi)}\\\hline
\rowcolor{LightCyan}
\textbf{User Input} & App installation, allow permissions & None required \\\hline
\rowcolor{LightCyan}
\textbf{Data Collection} & Direct from user device & WiFi infrastructure \\\hline
\rowcolor{LightCyan}
\textbf{Environment} &  Indoor and outdoor & Indoor and limited outdoor \\\hline
\rowcolor{LightCyan}
\textbf{Sensing Frequency} & App-dependent & Limited to target environment \\\hline
\rowcolor{LightCyan}
\textbf{Availability} &  BLE beacon instrumentation & WiFi network deployment  \\\hline
\textbf{Location Accuracy} & Fine-grained & Coarse-grained, AP-level \\\hline
\textbf{Power Consumption} & High impact from apps & Minimal impact from usual\\
\textbf{} & running in the background & WiFi signal scanning\\\hline
\textbf{Compatibility} &  App and device dependent & Device-dependent\\\hline
\textbf{Privacy/Security} & App-dependent (types of data, & Only WiFi network events are\\
\textbf{} & data collection practices) & collected. Identifiers are hashed.\\\hline
\end{tabular}}
\caption{Comparison between two forms of mobile sensing -- client vs. network-centric. Highlighted rows indicate the key adoption challenges to overcome.}
\label{tab:comparisonWifi}
\end{table}

As discussed in Section \ref{sec:usingWiFi}, the ability to use WiFi as a network-side positioning system is a key technology enabler for our research. Specifically, when a (user's) smartphone connects to an access point, it follows that its current location is within the vicinity of a nearby AP. Since enterprise WiFi networks log each connection made by mobile devices to each of their APs, this log can be analyzed to infer crowd movement patterns across campus and the occupancy levels in different buildings (based on the number of smartphones connecting to each AP).

Table \ref{tab:comparisonWifi} summarizes the comparison between Bluetooth/GPS-based sensing (\textit{client-centric}), more commonly adopted by present solutions, and WiFi-based sensing (\textit{network-centric}) technique. Our choice for sensing must be straightforward in overcoming the key adoption challenges faced by existing \textit{client-centric} applications. Specifically, the sensing approach must:
\begin{enumerate}
	\item \textbf{Take The Path Of Least Resistance:} Users need not install a dedicated mobile application or allow device permissions, such as what is required for Bluetooth or GPS-based sensing applications. Our technique can automatically discover and scan connected devices without explicit user interaction, making it much easier to deploy at scale.
	
	\item \textbf{Bypass User Device for Data Collection:} In the same way that no user action is required above, no user device will be accessed for data collection. Instead, WiFi network events, such as ``syslog'' and ``RTLS events'' will be collected directly from the WiFi infrastructure. Many enterprise networks already commonly use these events for IT security and performance monitoring.

	\item \textbf{Immediately Operate in Enclosed Areas:} Operability, especially in indoor spaces, remains a challenge by GPS-based techniques due to inaccurate satellite signals. Bluetooth-based applications heavily depend on setting up BLE beacons within the vicinity. In contrast, wireless networks are increasingly popular and available, specifically in work and education environments. Sensing location, however, is limited to users within the target environment.
\end{enumerate}

\label{sec:challenges}
\subsubsection{Challenge \#1: No WiFi Network} 
WiFi-based sensing is not without several challenges.
A key assumption of our work is \textit{ubiquitous network coverage}, in that WiFi coverage is present in all spaces where mobile sensing needs to be performed. With increasing and improving WiFi deployments on college campuses to meet student demands \cite{wifi62020}, it is reasonable to assume a near-universal WiFi coverage inside campus buildings. However, WiFi availability outdoors can vary with AP placements that are typically indoors. It is important to note that our goal is to monitor safety compliance following current public health guidelines, specifically in enclosed areas where \virus spread will pose more risk to occupants. 

\subsubsection{Challenge \#2: Disconnected Users}
The other key assumption is \textit{ubiquitous phone availability}, in that every campus user has a smartphone with them at all times. This directly implies that our sensing mechanism will overlook a user with no phone. Nevertheless, much research has argued for a high percentage of smartphone users, particularly on college campuses \cite{crompton2018use}. 

Additionally, smartphone users utilizing cellular data will present as \textit{unconnected mobile devices} in our approach. Despite widely available cellular data coverage, the WiFi network is still a preferred alternative provisioned for many online activities that demand low latency and high bandwidth networks. These activities include video/ music streaming and online gaming, which are enthusiastically engaged by students. A different source of unconnected mobile devices is visitors. While visitors of the university may not utilize the campus WiFi, their device remains visible to the WiFi network. This is because both iOS and Android devices are set to scan for available WiFi networks periodically (even though no connection is established), albeit with an anonymous MAC address. Logs of unconnected devices make precise monitoring challenging, but these records can still present as coarse-grained information approximating visitor occupants. Nonetheless, a natural course of handling crises such as \virus is clamping down on visitors (e.g., cancel open-houses and conferences) to prevent the spread outside the population. Our work proceeds to manage the occupancy and mobility of university residents who must resume their day-to-day work/school practices. Note that moving forward, we refer to unconnected devices as `unassociated devices.'

\subsubsection{Challenge \#3: Coarse-grained Location}
A final challenge is \textit{coarse-grained positioning} because WiFi networks can only yield positions at the granularity of AP location. While precise location (and proximity) information is important for digital contact tracing, this requirement is not necessary to achieve our goal. Our focus is on analyzing aggregated occupancy trends and the overall mobility patterns across the campus environment, to which coarse-grained position information is more than adequate.

\subsection{Privacy and Ethical Considerations}
\label{sec:privacyEthicalConsiderations}
Despite the noble intentions of combating \viruss, most digital solutions present the challenges of user privacy. WiFi-based sensing is no different, particularly since users' WiFi network data will automatically be collected and analyzed within the vicinity. Unlike most mobile sensing efforts for \viruss, we hope to determine WiFi-based sensing as a feasible way for institutions to monitor and maintain compliance with current public health standards -- this monitoring is accomplished at an aggregated scale and does not require identifying individual users occupying the facility. All identifiable information of users is anonymized and cannot be reverse-engineered. The main data source for our analysis, at present, is already being collected by network security administrators. Nonetheless, several privacy safeguards already exist to be put into practice. They are:
\begin{enumerate}
    \item \textbf{No access to user device:} Many users are likely unaware of the types and frequency of data being collected from their mobile devices. Indeed, no data will be transmitted directly from the user's device through WiFi sensing. Instead, our analysis will strictly utilize WiFi network data that is already collected by deployed wireless infrastructures. The frequency of data collection is constrained to the time users are within the vicinity. All identifiable information in the WiFi network data, particularly the MAC ID and username (if any), is anonymized using the SHA-2 hash.
    
    \item \textbf{Established network security protocol:} The National Institute of Standards and Technology (NIST) recommended that enterprise network security analysis is the best practice to build strong cybersecurity and protect an organization. Indeed, a variety of security logs, including WiFi network data, are already being used for auditing, supporting investigations, and identifying operational trends \cite{kent2006guide}. The need for this security protocol has led to laws and regulations compelling organizations to protect user privacy. Utilizing the same data source for \virus safety compliance will follow the same established security standards.
    
    \item \textbf{Emergency use authorization:} It is important to note that our analysis, purposed for institutional safety compliance, is presented at an aggregated scale and does not include pinpointed behaviors. However, this information may be deemed critical to further a contact tracing investigation. In such an event, emergency disclosures will be handled by an authorized official. We believe a formal operational protocol to assess risk variations must be in place before any information disclosure. Only when an individual is identified as at-risk can a public health case investigator obtain a de-anonymized copy of the information.
\end{enumerate}

\subsubsection*{Data Ethics}
All data used in this paper was obtained directly from the campus infrastructure and bounded by the computing agreements agreed to by each WiFi user when they received their WiFi credentials. These agreements allowed us to use their data for aggregate analysis as long as individual identifiers were not used. As such, every MAC address obtained from the WiFi infrastructure was hashed using a consistent 1-way hash function, and no specific user details (e.g., login IDs) were used. All of the analysis used by this paper focuses on large aggregates with no analysis of specific individuals performed.
\section{Data Collection and Methodology}
\label{sec:data}
Our approach is using WiFi data to infer \textit{occupancy} and \textit{mobility}.  In what follows, we state our assumptions and pre-processing steps prior to clarifying the key measures.

\subsection{Data Collection}
Our data sets were collected directly from the production WiFi networks of three different university campuses. Two of these campuses (\smu and \nuss) are in Singapore, and the last one (\amhersts) is in the Northeastern portion of the United States of America. Two campuses are full-sized residential campuses with over 200 buildings each and $\approx$40K to 50K students and staff, while the last university is a small non-residential establishment with $\approx$ 10,000 students and staff spread across seven buildings.

\begin{table}[h!]
 \scalebox{0.8}{
 \centering
 \begin{tabular}{|c|c|c|c|c|} \hline
 \textbf{Campus} & \textbf{No. Buildings} & \textbf{No. Students} & 
 \textbf{No. Staff} & \textbf{No. APs} \\ \hline
Singapore Management University -- \smu  & 7 & $\approx$9,000 & $\approx$1,000 &  $\approx$800 \\ 
National University of Singapore -- \nus  & $\approx$240 & $\approx$40,000 & $\approx$10,000 & $\approx$13,000  \\ 
University of Massachusetts Amherst -- \amherst & $\approx$230 & $\approx$30,000 & $\approx$8,000 & $\approx$5,500  \\ 
\hline
 \end{tabular}}
 \caption{Details of each campus studied}
 \label{tab:data}
\end{table}

All three universities run Aruba-equipment supplied WiFi networks, with one university also running a Cisco-equipment supplied WiFi network in addition to an Aruba network. For the Aruba networks, we pulled the WiFi data directly from the infrastructure using either real-time location services (RTLS) APIs~\cite{ArubaRTLS} or by reading the system logs directly. For the Cisco network, we pulled WiFi data directly from the network using the Cisco Connected Devices (CMX) Location API v3~\cite{CiscoCMX} (recently rebranded as Cisco DNA Spaces). In all cases, we obtained the following information:  for all associated WiFi devices, the timestamp when the associated device was seen, the BSSID of the Access Point (AP) that saw the device, and the hashed client MAC address of the associated device. For two of the campuses, we can also obtain the same information (time seen, BSSID that saw the device, and hashed client MAC) for unassociated devices as well -- i.e., devices with WiFi on that are just scanning. 

We have associated device data from Feb 2020 onwards for all three campuses, allowing us to clearly view campus occupancy and mobility patterns across campus before, during, and after COVID-19 related measures were implemented. Table~\ref{tab:data} provides details of each of the three campuses as well as the data collected.

\subsection{Using WiFi Positioning System}
\label{sec:usingWiFi}
\subsubsection{Key Assumptions}
As described in Section \ref{sec:whywifi}, our analysis is based on the key assumption that most of our users frequently utilize campus WiFi on their own smartphones. A separate analysis on our US campus, \amhersts, reported approximately 90\% university residents carrying a smartphone with them at all times; specifically, our WiFi network events revealed 30,084 users comprising 24,791 student users and 5293 staff/faculty. The remaining 10\% neither owned a smartphone nor chose to use the campus WiFi. Like \virus digital contact-tracing apps, a critical obstacle in enabling an effective crowd monitoring solution is mass user adoption. While our key assumption naturally disregards a small percentage of users, the approach can achieve a critical mass of data and ascertain some occupancy and mobility measures among smartphone users. In what follows, we describe how different mobile users can be tracked.

\subsubsection{Pre-processing}
We maintain similar pre-processing steps to extract information on human mobility for all campuses. For all WiFi enterprise networks deployed in the three universities, each AP on the network internally keeps a log of \textit{association}, \textit{disassociation}, \textit{authorization}, and \textit{un-authorization} activities with the devices that help us compute occupancy at various locations. An AP system log or ``syslog'' comprises a sequence of timestamped events. Each of these events follows the format:

\begin{verbatim}
hh:mm:ss <controller_name> <process_id> <event_subtype> <MAC_addr> <event_body>
\end{verbatim}

For each event in syslog, an event\_subtype representing the network event type is specified. This code can be recorded as six event types. They are \textit{association}, \textit{disassociation}, \textit{re-association}, \textit{authentication}, \textit{deauthentication}, and \textit{drift} events. Based on time (i.e., timestamp) and AP location, defined by controller\_name and event\_body, we can compute connection sessions per device across all APs \cite{trivedi2021wifitrace}. For each device, we create a timestamp indexed sequence of sessions to acquire device trajectories. Combining these data sets helps us to characteristically produce user-profiles (e.g., authenticated users as university residents) and activities (i.e., occupancy at a dedicated location or transitioning between locations). Note that \textit{MAC\_addr} is anonymized, as described in Section \ref{sec:privacyEthicalConsiderations}.

\subsubsection{On-campus Deployment} 
\begin{figure}[h!]
\includegraphics[width=.8\linewidth]{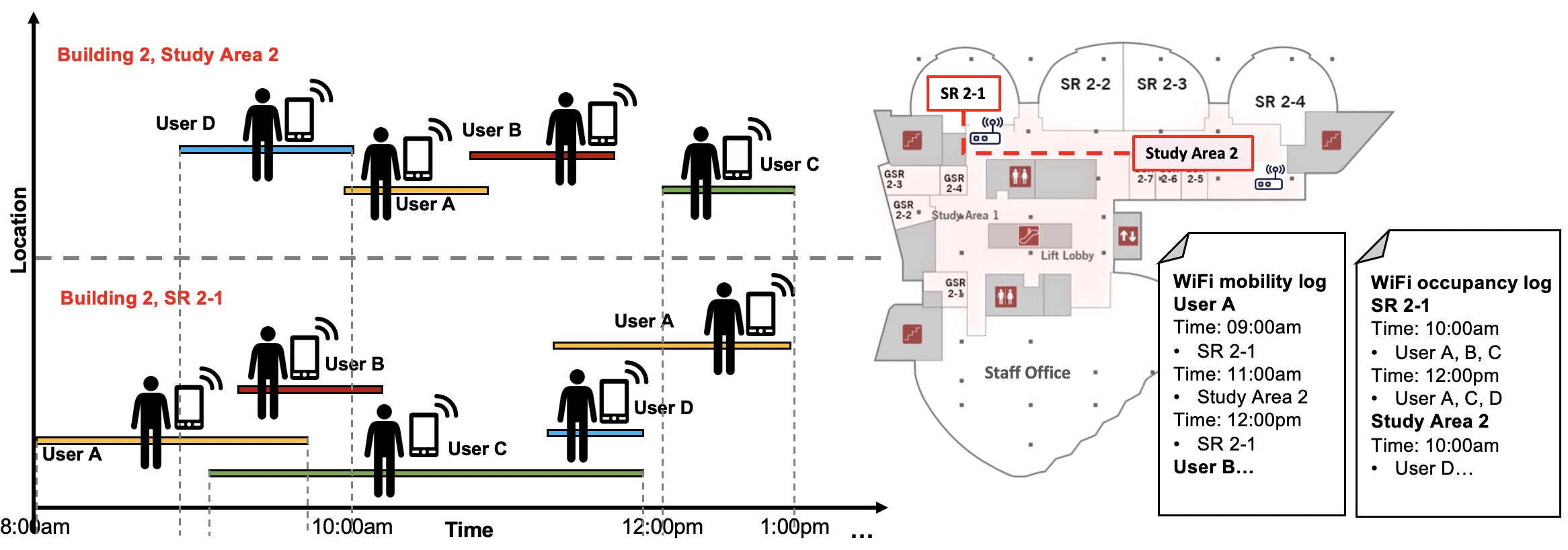}
\caption{Infer occupancy and mobility based on smartphone WiFi connectivity to the nearest AP at floor-level.}
\label{fig:wifiData}
\end{figure}

We illustrate how network events (when the smartphone is connected to the campus WiFi) will produce time-based traces in Figure \ref{fig:wifiData}. A deployment of an enterprise WiFi network consists of many access points (AP), optimally spread across buildings and floors. 

When a user first utilizes the campus WiFi network on their own smartphone, \textit{authentication} will occur (through their student ID and password), resulting in an \textit{authentication} and \textit{deauthentication} log messages. Note, however, student identification is hashed to preserve user privacy (see Section \ref{sec:privacyEthicalConsiderations}). Simultaneously, this action triggers the device to connect with the nearest AP, generating \textit{association} log messages. The device continues to stay connected to the AP until the user moves; in this case, the connection switches to the next nearest AP to where the user is now situated. Accordingly, \textit{disassociation},  \textit{re-association}, and \textit{drift} log messages will be generated when the user's device moves out of range, reconnects from `sleeping.' Throughout the whole time, the user maintains the same network connection to the campus WiFi, only switching APs.

In Section \ref{sec:challenges}, we discuss the challenges of accounting for unconnected devices, specifically devices that remain connected to their personal cellular network. This group of users will still be accounted for as long as their devices are scanning for WiFi network (i.e., unassociated devices). It is important to note, however, our main analysis only consists of associated devices, representing university staff and students.

There are two other sources of error from utilizing this technique. First, a smartphone may maintain its existing connection to an AP due to good signal strength even though the user has moved. This error makes no impact on occupancy since we are counting on per floor/building basis, but will produce small errors on mobility. Second, a user may connect multiple devices (e.g. smartphone, laptop, wearable) to an AP at once. However, duplicated events can be removed \cite{trivedi2021wifitrace}, while device types can be further filtered to only collect one specific type if needed. This methodology has been validated by others \cite{lu2016robust,trivedi2021wifitrace,hwang2017process,zakaria2019stressmon} and used for similar applications of mall analytic and queue management \cite{jayarajah2016livelabs}.

\subsection{Inferring Occupancy and Mobility Using WiFi Logs}
\label{sec:inferMeasure}
A campus network comprises several user types such as faculty, staff, students, on-campus student residents, and visitors. The syslog authentication event consists of login types, thus helps differentiate a faculty/staff from a student. We further subdivide user groups based on the following rules. Students who spend more than 5 hours at an on-campus residential dormitory will be classified as on-campus student residents. Visitors are classified in several ways. First, records of anonymous MAC addresses are regarded as visitors (see Section \ref{sec:challenges}). Second, users recorded with only a one-time login or devices with only several days of login over the course of the semester are most likely visitors. This heuristic is necessary to filter out users attending one-time events such as hack-a-thons, open houses, and conferences held on campus.

In Sections \ref{sec:analysis_sin} and \ref{sec:analysis_us}, respectively, our analyses will include reports of location occupancy at three-level granularity: area, floor, or building occupancy based on a collection of WiFi access points (i.e., AP location). For example, a large room such as a common dining hall can have more than one WiFi AP, while one AP can be at the intersection of different rooms. The coarse-grained information based on AP location amounts to inaccuracies in determining room-level information, especially if the room is small. Thus, an area can consist of a collection of small rooms (e.g., see Figure \ref{fig:wifiData}, GSR2-2, 2-3, 2-4 is regarded as a single area) or a singular large room (e.g., Figure \ref{fig:wifiData}, SR2-1). Accordingly, floor occupancy is the total of all areas on each floor, and building occupancy is occupancy on all floors in each building. We show in Section \ref{sec:sin_occupancy} the changes in occupancy rate over different \virus phases, specifically for these areas.

\begin{figure}[h!]
\includegraphics[width=0.9\textwidth]{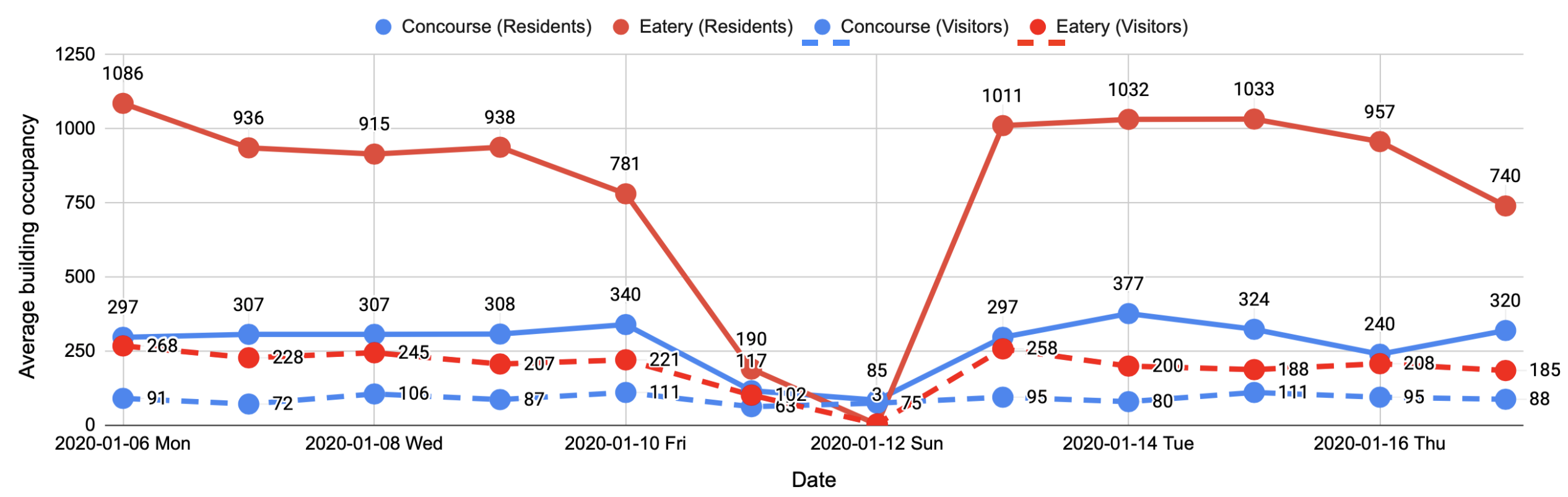}
\caption{Daily occupancy trends for two common areas, the concourse (blue) and the main eatery (red), based on associated devices (line) and unassociated devices (dashed line) on school days prior to the \virus outbreak. Both areas are located on the same floor within a building at \smu campus. \label{fig:assocDeassocEvents}}
\end{figure}

\subsubsection{Definition: Occupancy}
We determine \textit{occupancy} as the average number of people in a dedicated area of single building floor, as shown in Figure \ref{fig:assocDeassocEvents}. By taking the average occupancy at building level per day, our features can avoid the problem of missing data, common in longitudinal data collection procedures. The number of people is defined by unique device counts (smartphones) logged in our WiFi network records. The groups of people can also be known by identifying associated devices (for university residents) and unassociated devices (for visitors). For example, the \smus concourse and main dining hall record approximately 100-200 visitors pre-COVID (January 6 to January 16, 2020). This separation is necessary, especially since our universities are open campuses, providing pedestrian access to the public. Such information
can aid institutions in identifying what percentage of occupants are university residents should they backtrack along the flagged path and areas vulnerable to exposure.

\begin{figure}[h!]
\includegraphics[width=0.7\textwidth]{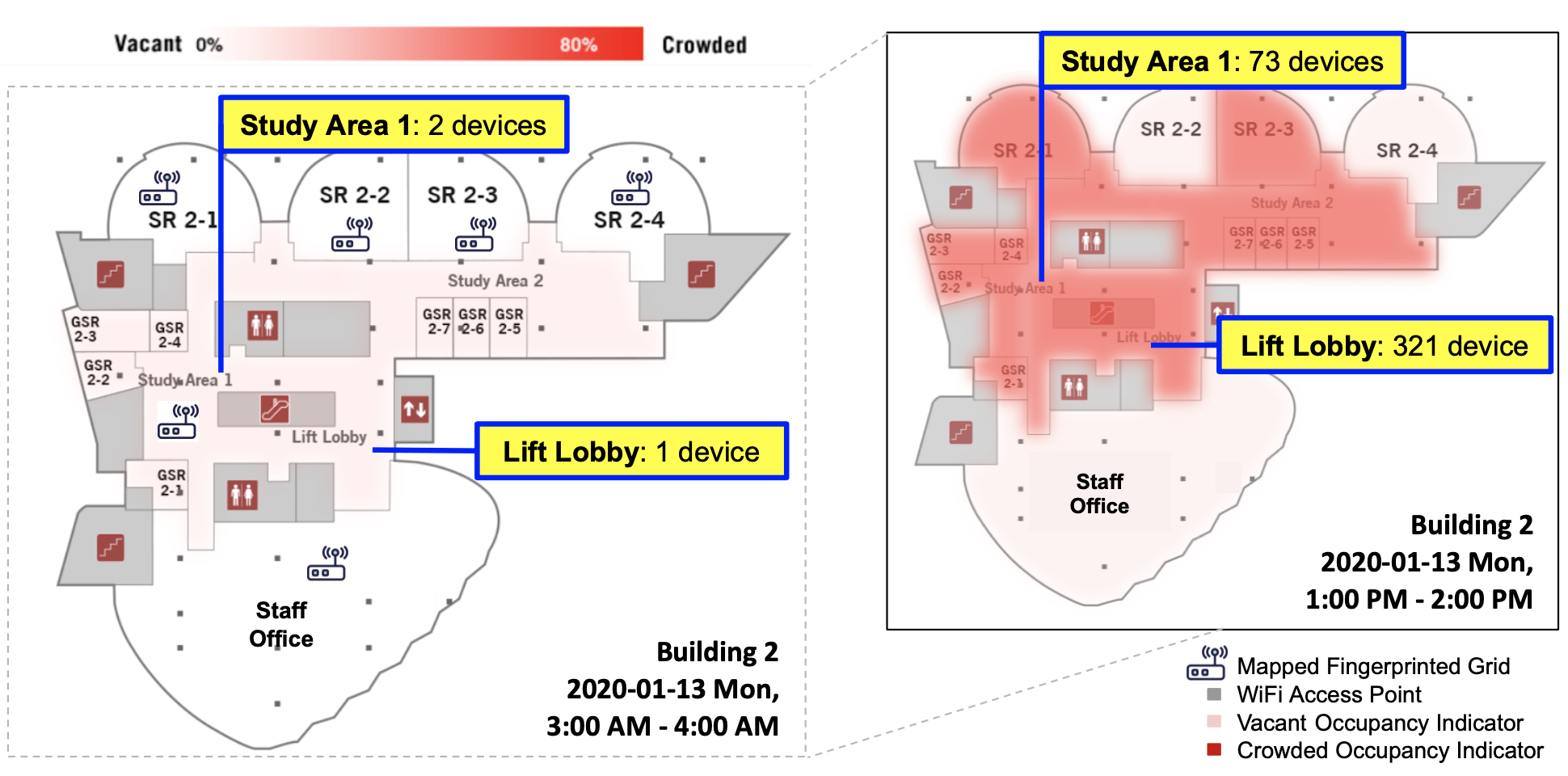}
\caption{Heat map of a floor in Building 2 of \smu campus. Left shows the occupancy in the early morning, while right map shows occupancy during a regular school hour.  \label{fig:buildOcc}}
\end{figure}

Examining our data from a different perspective, occupancy reveals cues about clusters on different areas of each floor building per hour. Figure \ref{fig:buildOcc} depicts a floor map in Building 2 at two time periods, one in the early morning and the other during school hours. By directly pulling a collection of network event logs from APs surrounding Study Area 1, we can determine 2 unique devices connected between 3:00 AM - 4:00 AM, indicating low building occupancy. In contrast, logs between 1:00 PM - 2:00 PM on a Monday show the expected high occupancy (e.g., 73 device connections at Study Area 1) throughout the hour.

\begin{figure}[h]
\begin{center}
\begin{tabular}{cc}
    \includegraphics[width=1.3in,angle=90]{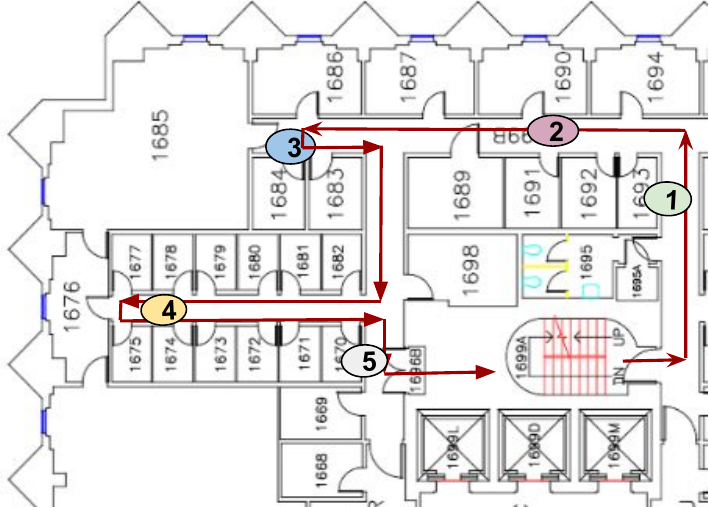}&
    \includegraphics[width=2in]{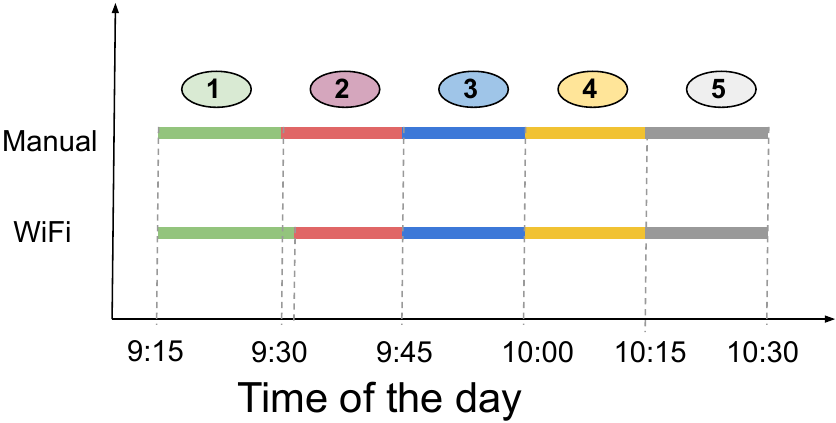}\\
    (a) & (b)
\end{tabular}
\end{center}
\caption{(a) Floor map with AP locations within the \amherst campus (b) Temporal lag between WiFi extracted trajectory and ground truth manual log}
\vspace{-0.05in}
\label{fig:US_mobility}
\end{figure}

\subsubsection{Definition: Mobility}
\textit{Mobility} is defined as the average number of areas visited by a user per hour; once again, an area is defined by building level, measured for each floor comprising a collection of rooms on that ground. Figure \ref{fig:US_mobility} shows the floor map of a campus building in \amhersts, comprising individual office units, classrooms, and open spaces. As a user moves around the floor through AP 1 to 5 - indicated by the red arrow line in figure \ref{fig:US_mobility}(a), we observe changing network events from the device, generating \textit{association} and \textit{disassociation} events across different APs. The WiFi trajectory from one AP to another allows us to determine a user being `stationary' or `transitioning' between areas on the same floor. Figure \ref{fig:US_mobility}(b) shows the temporal lag recorded for one user from switching to AP2 as he transitions. While temporal lags are likely to happen when APs are positioned close to one another, these inaccuracies will not affect our analysis in capturing building-level transitions. 

\begin{figure}[h!]
\includegraphics[width=0.8\textwidth]{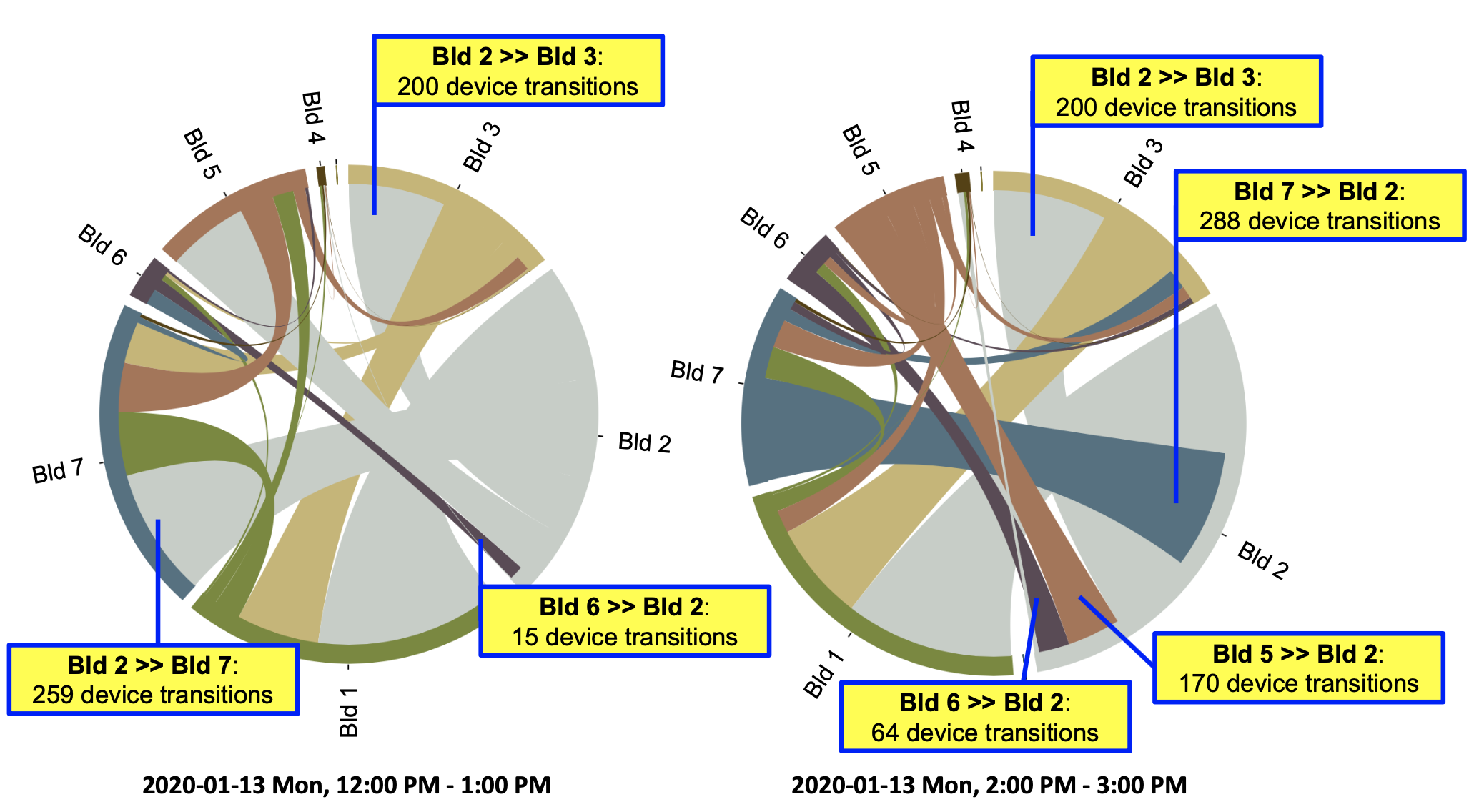}
\caption{Chord diagram representing the number of devices recorded to move from one building to another at different times of the day. The diagram displays all seven buildings within the \smu campus. \label{fig:mobilityBuilding}}
\end{figure}

In the context of \viruss, one of the concerns is identifying possible transmission routes within and across the institution's buildings. Figure \ref{fig:mobilityBuilding} comprises two chord diagrams representing the number of users recorded moving to and fro across seven different buildings within the \smu vicinity. The left diagram illustrates much traveling \textit{into} Building 7 from all other buildings during lunch hour, 12:00 PM to 1:00 PM. For example, 259 device transitions were made from Building 2 alone. In contrast, we observe more transitions distributed across buildings between 2:00 PM to 3:00 PM. For example, occupants recorded to be in Building 2 during that hour had visited Building 7 (288), 5 (170), and 6 (64 devices) before their transition. With some buildings being a single commonplace for occupants across campus to gather at different times of the day, this information can aid institutions to implement \virus policies, for example, deploying social distancing support for regulating crowd control.

Conclusively, these statistics gathered from WiFi network logs can provide occupancy and mobility information, allowing us to monitor high foot traffic areas within the universities. In the next section, we investigate the adequacy of WiFi data in revealing the spectrum of changes in occupancy and mobility when the \virus safety measures were enforced.
\section{Mobility and Occupancy Analysis for Singapore Universities}
\label{sec:analysis_sin}

\begin{table}[t!]
\centering
\scalebox{.8}{
\begin{tabular}{llp{.6\textwidth}} \hline
\textbf{Singapore Phases} & \textbf{Date Started} & \textbf{Safety Measures} \\ \hline
Pre \virus  & Before  & Awareness on personal hygiene \\
($P_S0$) & 19 Feb, 2020 & \smu moves some classes online \\ \hline
Phase 1 ($P_S1$) & 19 Feb, 2020 & National threat level raised to orange~\cite{Dorscon_Orange} \\
& & 14-days home quarantine enforcing if returning from China~\cite{China_Stay_Home} \\ 
& & All core curriculum moved to online learning for \nus \\ 
& & In-class mid-term assessment cancelled for \nus \\
& & both \smu and \nus implement a 1 meter safe distancing policy \\
& & \smu closes all sports facilities \\
& & Classes >= 50 students moved to online learning for \smu and \nus \\ \hline 
Phase 2 ($P_S2$) & 22 Mar, 2020 &  All travel cancelled unless mandatory \\ 
& &  All visitors to Singapore issued 14 day Stay Home order~\cite{All_Stay_Home} \\
& &  \smu enforcing A/B shifts where all students \& staff \\ 
& &  must alternate being offsite every other week \\ \hline
Phase 3 ($P_S3$) & 3 Apr, 2020 & Full shift to online learning for all schools at all levels~\cite{MOE_Online_Learning} \\
& & All exams moved online at \smu and \nus \\
& & Pass / Fail option offered to students at \smu and \nus \\
& & \smu only allowing key personal on campus \\ 
& &  \nus allowing most employers to work from home\\ \hline
Phase 4 & 7 Apr, 2020 & Full country-wide stay at home quarantine~\cite{Circuit_Breaker} \\ 
  ($P_S4$, ongoing) & & Nobody allowed on campus for \smus. All buildings closed. \\ 
& & Only approved students allowed to stay in dorms at \nus \\
& & Approved students allowed to travel to \nus to study \\
& & Measures extended until Jun 2020~\cite{Circuit_Breaker_Extended} \\ \hline
\end{tabular}}
\caption{Five phases of \virus related safety measure enacted in Singapore and at \smu and \nus}
 \label{tab:covid_dates_sg}
\end{table}

We examine the changes in occupancy density and movements on-campus to determine the effectiveness of various safety measures put in place at significant times points of \viruss. These time points are summarised in Table~\ref{tab:covid_dates_sg}.

\subsection{Overall Control Policy}

Singapore was first alerted of `severe pneumonia' cases in Wuhan city on 2nd January, 2020. From that point on, Singapore's Ministry of Health (MOH) has mandated a series of escalating control policies to prevent high infection rates of \virus while minimizing significant disruptions to the daily routines of its residents. The first case of \virus in Singapore was reported on 22nd January, and more cases started appearing over the next few days. As shown in the subsequent phases ($P_S1$ to $P_S4$, lockdown\footnote{The stringency of ‘lockdown’ in Singapore was relatively modest compared to those in China, Italy, and Australia, where people could not dwell or travel beyond their immediate neighborhood.}), Singapore's MOH took increasingly strong measures to contain the spread of infection. These measures included mandatory stay-at-home quarantine orders for visitors, moving all academic programs online, to enforcing country-wide stay-at-home orders. Additionally, numerous facilities across Singapore were re-purposed as quarantine centres. For this analysis, several dormitory blocks at the \nus campus were re-purposed for this use in early May 2020~\cite{SIN_NUS_Workers}.

\subsection {Impact of Different Policies on Occupancy}
\label{sec:sin_occupancy}

\begin{figure}[h!]
    \centering
    \includegraphics[width=0.8\textwidth]{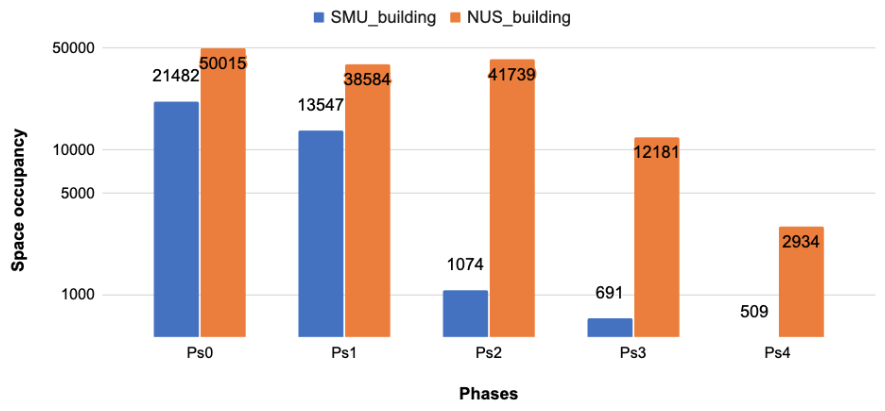}
    \caption{Building-level occupancy for one school building for each university, \smu (blue) and \nus (orange), plotted from $P_S0$ to $P_S4$.}
    \label{fig:schoolbuilding}
\end{figure}

\begin{figure}[h!]
    \centering
    \includegraphics[width=0.495\textwidth]{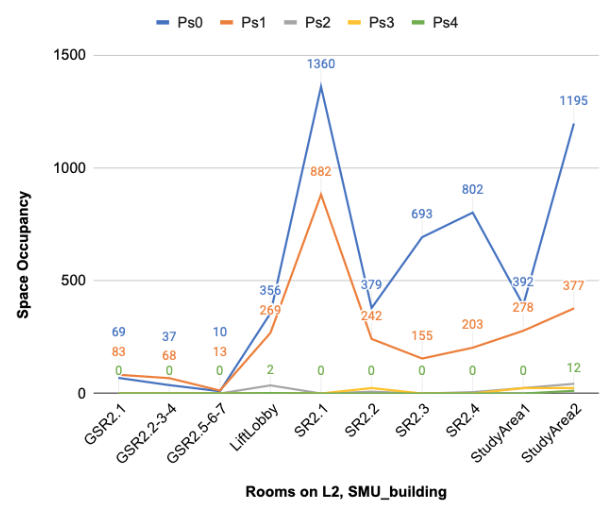}
    \includegraphics[width=0.495\textwidth]{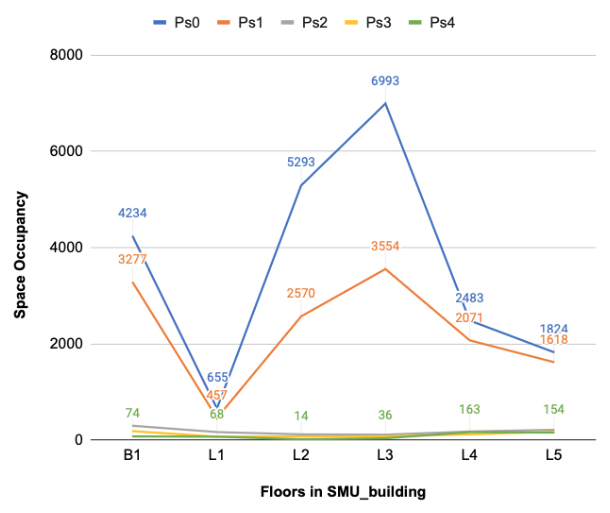}
    \caption{The percentage change in area-level occupancy within different rooms (left) and levels (right) of \smu buildings, plotted from pre \virus time ($P_S0$) to current time ($P_S4$).}
    \label{fig:roomfloorSMU}
\end{figure}

We first computed the drop in campus occupancy in Singapore as each phase of \virus related policies were enacted. Figure~\ref{fig:schoolbuilding} plots the unique device counts for one building within the Singapore universities, \smu and \nuss, over the \virus phases, $P_S0$ to $P_S4$. We observed there was a more than 90\% percentage drop from phase $P_S0$ to $P_S3$ and beyond for \smu when the university implemented an almost full work-from-home policy ($P_S3$) followed by the nationwide lockdown ($P_S4$). Despite taking the same set of measures, \nus was successful in reducing \textit{occupancy} to only 75\% at $P_S3$, and 98\% by the lockdown. For \smus, the drop was almost 100\% by Phase 4 ($P_S4$) as nobody, except for security staff, was allowed onto campus whereas \nus still allowed a few thousand students to stay in the dorms. Figure \ref{fig:roomfloorSMU} charts the occupancy rate at room-level and floor-level for \smus. A drastic drop in occupancy can be observed as soon as in Phase 1 ($P_S1$) when large classes were shifted to online learning—the space utilization specifically for the seminar rooms (SR) 2.1 to 2.4, which are regularly used for holding classes, decreased by more than 50\% on average. Additionally, overall occupancy declined the most for levels 2 and 3, consisting mostly of seminar rooms. While no one was not allowed to work on campus during $P_S4$, clearances were granted and arranged for personnel to bring their workstations home. Note that the occupancy of 12 at study area 2 during $P_S4$ did not last for more than 15 minutes.

\begin{figure}[h!]
    \centering
    \includegraphics[width=0.495\textwidth]{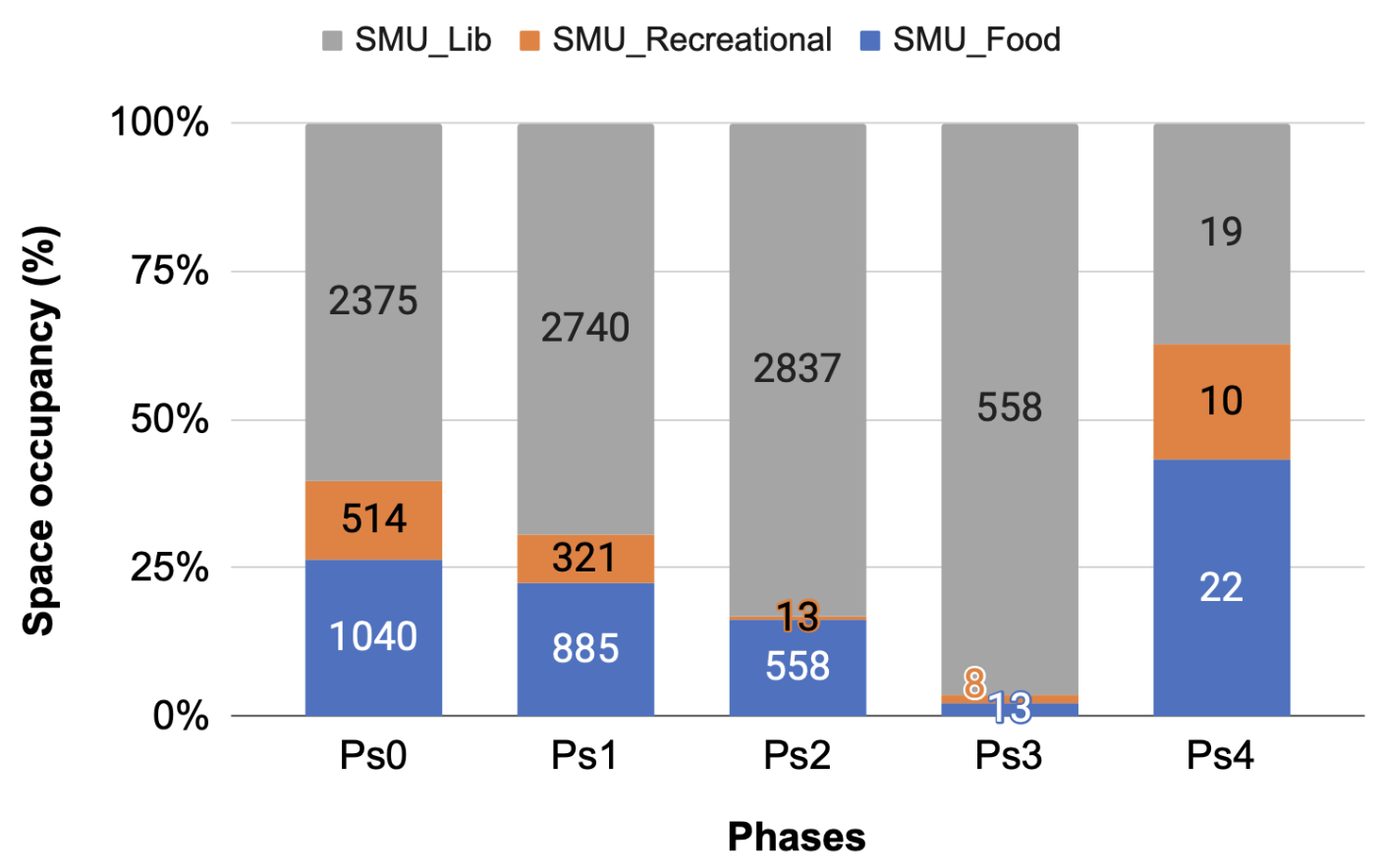}
    \includegraphics[width=0.495\textwidth]{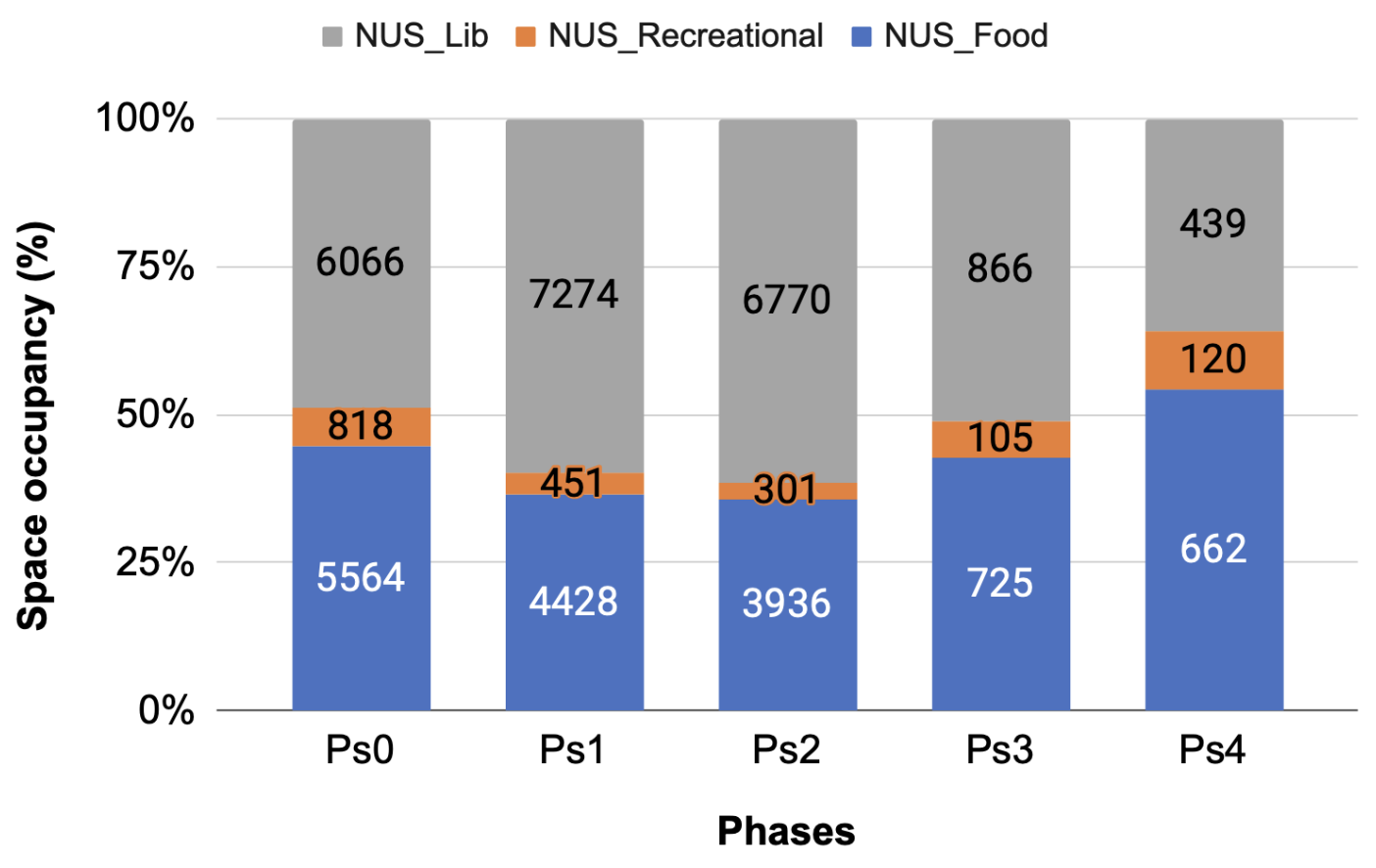}
    \caption{The percentage change in area-level occupancy within different \smu (left) and \nus (right) buildings, plotted from pre \virus time ($P_S0$) to current time ($P_S4$).}
    \label{fig:schoolAreas}
\end{figure}

To understand this change of occupancy in more detail, Figure~\ref{fig:schoolAreas} shows the percentage change in space occupancy for different types of areas located at three buildings per campus. These areas are dedicated to three activities: recreational, dining, and studying. We considered only the indoor gym area within a building to represent recreational activity, the only common dining hall area within a building to represent dining activity, and multiple study areas on all floors within the library building to represent study activity. The values for \smu are shown in the left figure, while the values for \nus are shown in the right figure. The absolute count for each percentage is listed inside each area (e.g. The absolute count for \smus\_Food in $P_S4$ is 22 people comprising about 40\% of the total occupancy).

There were differences in the space utilization between the two campuses. For example, \smu  closed all recreational facilities in $P_S2$ onwards, and this is reflected in the noticeable percentage drop. The on-campus dining facilities were also mostly closed from $P_S2$ onwards.

As \nus has a large number of students staying on campus in dorms (\smu does not have dorms on campus), even in $P_S4$, the occupancy of recreational spaces was relatively high (and higher percentage-wise than earlier phases). This result is likely because students staying in dorms did not want to stay in their rooms all day long and decided to go out to these recreational spaces (which is technically a violation of the quarantine rules in effect). 

Such occurrences raise concerns that these recreational spaces would have larger than optimal occupancy densities and undesirable mixing students from different dorms that would compromise measures designed to contain the spread of infection. 

\subsection{Impact of Various Policies on Mobility}
\label{sec:sin_mobility}
\begin{figure}[h!]
    \centering
    \includegraphics[width=0.8\textwidth]{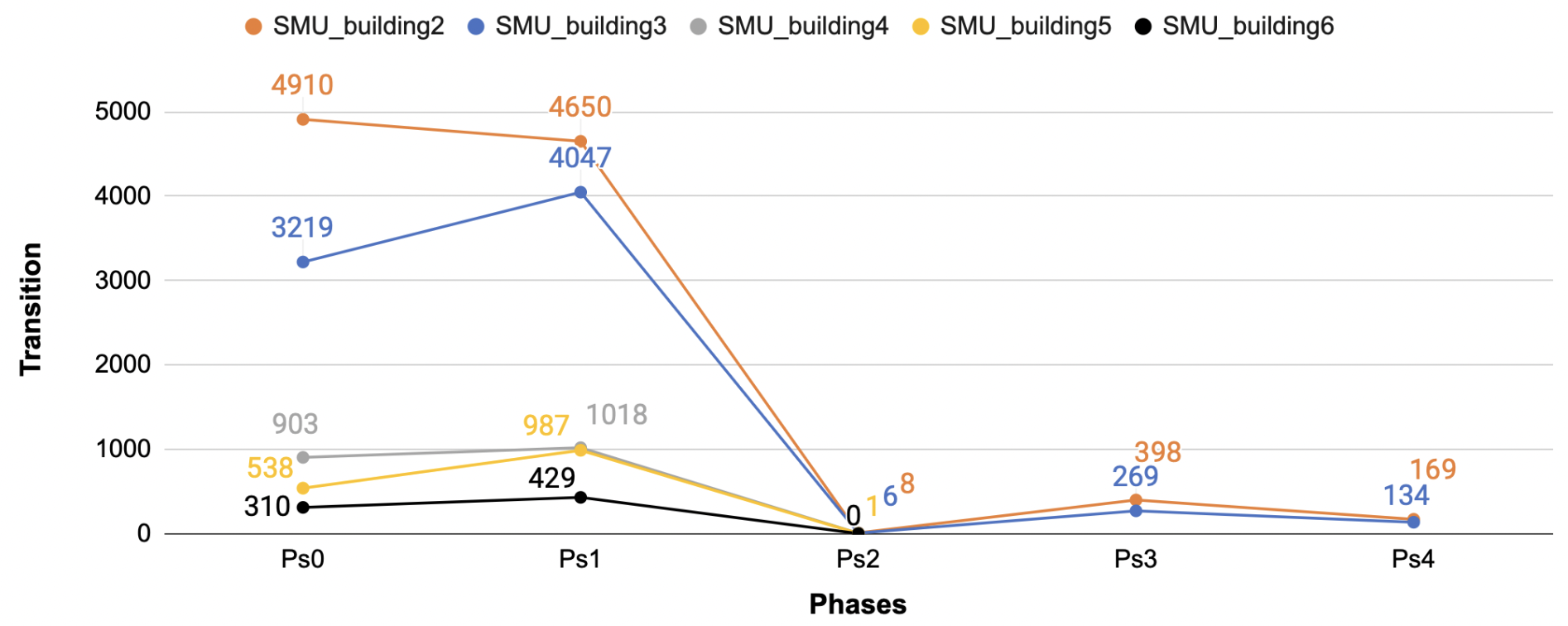}
    \caption{Device transitions originating from \smus\_building1 to five other buildings. The patterns of mobility remained the same on the overall despite noticeable reduction in the number of transitions over time.}
    \label{fig:SIStransition}
\end{figure}

Next, we sought to determine how various social mobility control measures influenced mobility patterns across both universities. First, looking at \smus, Figure~\ref{fig:SIStransition} charts the average transitions made on a per-building level over a day in each phase, from one building (called \smus\_building1), to five other \smu buildings. The transition count indicates the number of times a person moved from \smus\_building1 to the indicated building on that day. 

The results showed an expected decrease in the overall numbers as each phase was introduced, beginning with the university's decision to conduct online learning for its students at phase $P_S1$. In phase $P_S2$, \smu introduced full A/B schedules where only half the student and staff population could physically be on campus at any one time. This step reduced the overall occupancy (as shown in Figure~\ref{fig:schoolbuilding} and decreasing Transition count for $P_S2$ in Figure~\ref{fig:SIStransition}). 

\begin{figure}[h!]
    \centering
    \includegraphics[width=0.8\textwidth]{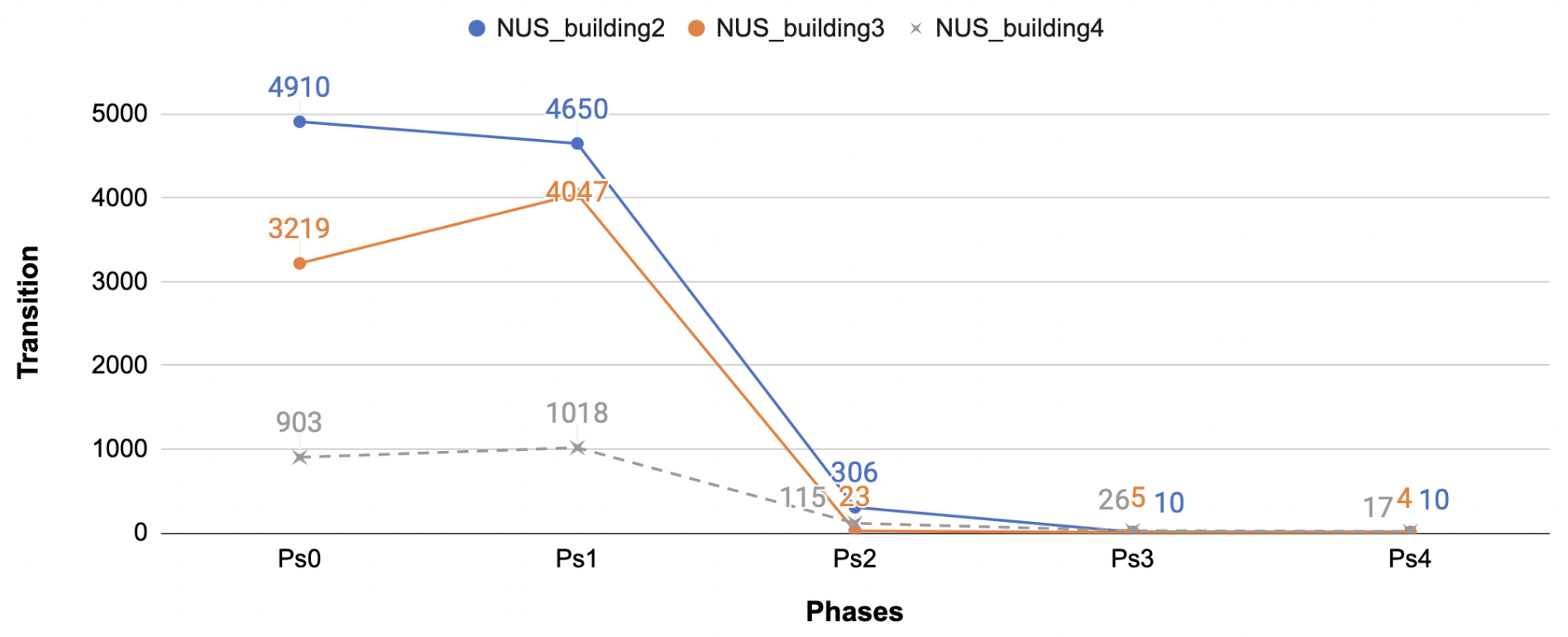}
    \caption{Device transitions originating from \nuss\_building1 to three other academic buildings over different phases. Overall, occupancy rate and mobility rate  are evidently less.}
    \label{fig:NUStransition}
\end{figure}

The actual mobility rate has also decreased for \nus due to decreased occupancy on campus. However, the mobility rate for each person on campus remained the same - this is understandable as the work required them to visit the same set of buildings they had previously. \smu reduced the campus occupancy to almost 0 in Phases $P_S3$ onwards, and this naturally reduced the mobility rate (Figure~\ref{fig:SIStransition}). 

We next looked at data from \nus to understand if these changes in mobility patterns were consistent. Figure~\ref{fig:NUStransition} shows the absolute number of transitions (numbers within each bar) along with the percentage of transitions from one \nus academic building to three other academic buildings. Similar to \smus, even though the total occupancy of the campus decreased due to the measures enacted in $P_S2$, the mobility rate (amongst those staff and students still on campus) remained high until more complete lockdown policies were enacted in $P_S3$ onwards.

\subsection{On-campus Living}
\label{sec:dorm_living}
The previous analysis focused on academic buildings. However, \nuss, unlike \smus, has a significant student population still residing in on-campus dormitories even during the complete lockdown phase $P_S4$. We dug deeper to understand the behavior of students living in these \nus dorms. 

\begin{figure}[h]
    \centering
    \includegraphics[width=.9\textwidth]{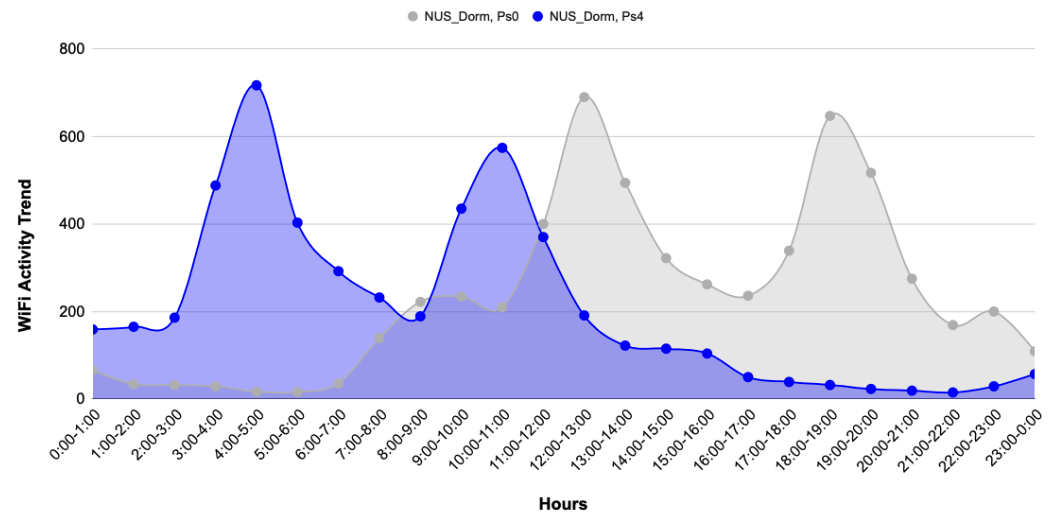}
    \caption{The hourly WiFi activity trend for \ndorm plotted before ($P_S0$) and during full lockdown ($P_S4$).}
    \label{fig:dormTrendDiff}
\end{figure}

\begin{figure}[t]
    \centering
    \includegraphics[width=0.8\textwidth]{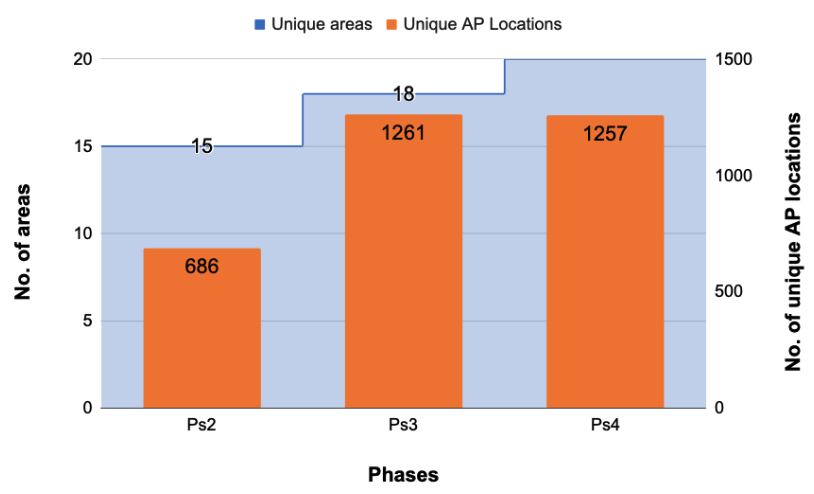}
    \caption{Transitions of 200 sampled \ndorm occupants to different (neighbourhood) area (e.g., another dorm) as a result of their phones being connected to unique AP locations within \nuss.}
    \label{fig:dormUserHistory}
\end{figure}

Figure \ref{fig:dormTrendDiff} plots the daily WiFi activity trend of one dormitory location, \ndorms, over phases $P_S0$ and $P_S4$. Overall, we observed the same daily activity levels (approximately 700) for this dorm across all phases -- indicating that the dorm population occupancy had not changed between phases. Instead, we observe the full lockdown at $P_S4$ resulting in the reversal of the WiFi activity trend with decreasing connections during the day and active WiFi utilization over the night, indicating user activeness at night.

Our analysis of mobility patterns amongst students staying in the dorms revealed some interesting findings. In particular, we found that even during lockdown periods ($P_S4$), a significant number of people were moving actively across campus -- which is technically a violation of the lockdown rules. Figure \ref{fig:dormUserHistory} presents the mobility rate of 200 randomly selected individuals staying in \ndorm to unique areas and AP locations visited within the campus vicinity. We found that their mobility rate increased during $P_S3$ and $P_S4$ compared to $P_S2$. Each occupant was making at least three transitions on average in phase $P_S2$, which doubled to about six transitions in $P_S3$ onwards. 

While the number of AP locations that individuals' devices were connected to seemed surprisingly high at first glance, these APs remained in neighboring areas of the users' residences. Specifically, individuals moved between areas in the same dorm to visit the dining, recreational facilities, or the bus stops more often (to head to other dining and grocery areas). These were all shorter length transitions compared to earlier phases -- in earlier phases, the transitions had long durations as the individuals were going to academic buildings for coursework reasons. 

\subsection{Main Takeaways}

From the analysis of \smu and \nuss, the main takeaway we derive is that policies that allow telecommuting and split-team load balancing are excellent for reducing the people density on campus. However, the staff and students that do work on campus are more likely to continue visiting the same set of places they utilized, thus, can lead to serious issues if an outbreak occurs -- as the \virus can be easily spread to all the other people visiting those areas. Thus, to avoid uncontrolled outbreaks, it may be necessary to limit the mobility of individuals and the only policy that was successful at this (from the many policies that were tried in Singapore) is a full lockdown where everyone is given stay-at-home orders. 

In addition, the mobility analysis of dorm occupants at \nus suggests that reducing mobility will require providing everything occupants need at their premises itself. Otherwise, the mobility rate could go up (even if the actual length of the transitions are shorter in duration) as individuals travel to other places to procure food and other essential items needed during a lockdown.
\section{Mobility and Occupancy Analysis for US University}
\label{sec:analysis_us}

Unlike Singapore, the US state our campus (\amhersts) was located in only had a single response -- the state went from business as usual to a stay-at-home order with shutdowns of many businesses~\cite{US_Response}\footnote{Note: for anonymity reasons, we cite an article listing all the states that have effected a similar policy.}.

\begin{table}[h!]
\centering
\scalebox{.8}{
\begin{tabular}{llp{.5\textwidth}} \hline
\textbf{US State Phases} & \textbf{Date Started} & \textbf{Safety Measures} \\ \hline
 Pre \virus ($P_U0$) & 29 Feb, 2020 & No Policy. Business as Usual. \\ \hline
 Phase 1 ($P_U1$) & 20 Mar, 2020 & Stay at Home State Wide Order~\cite{US_Response} \\ 
 & & No classes at \amherst \\ 
 & & Dorms cleared at \amherst except in special cases \\ \hline
\end{tabular}}
\caption{Dates corresponding to the safety measures for \virus in \amhersts}
\label{tab:covid_dates_us}
\end{table}

\begin{figure}[h!]
    \centering
    \includegraphics[width=0.65\textwidth]{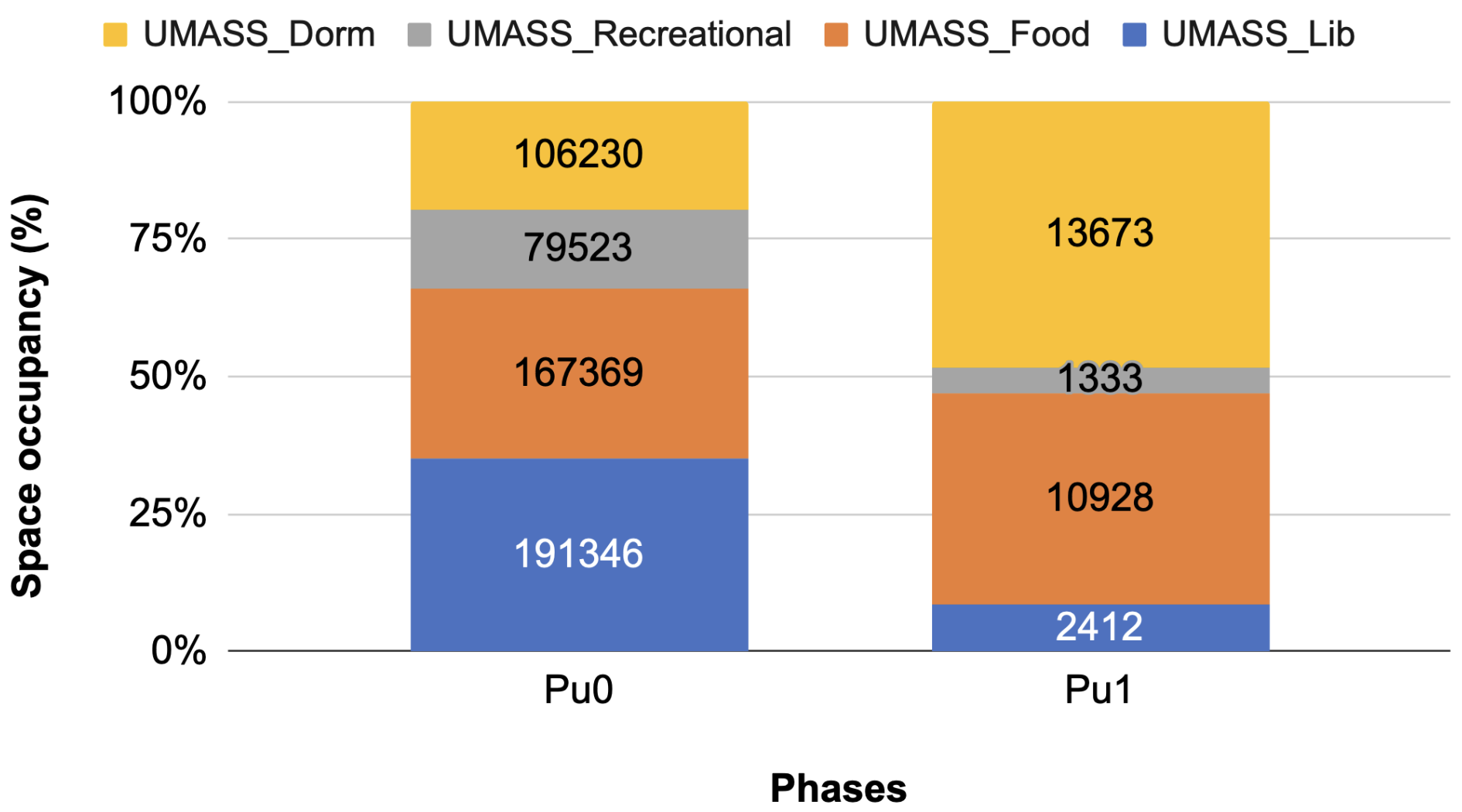}
    \caption{Space occupancy within \amherst for both $P_U0$ and $P_U1$ -- broken down by usage type}
    \label{fig:umassOverall}
\end{figure}

We observed similar occupancy trends, compared to \smu and \nuss, as \amherst transitioned into a lockdown. Figure~\ref{fig:umassOverall} plots the total number of unique devices detected for different area types over ten days for each of the two phases. The areas picked were ``Recreational'' (e.g. gym), Dorm (e.g. on-campus dormitory housing), ``Lib'' (campus libraries), and ``Food'' (e.g. food courts). Overall, we observed a more than 90\% decrease in occupancy between phases $P_U0$ (business as usual) $P_U1$ (full stay-at-home orders). In addition, we observed that the quarantine policies had naturally shifted the occupancy rates with the dormitories becoming the most occupied locations during $P_U1$, and consequently reducing the occupancy of the previously busy library areas.

\begin{figure}[h]
    \centering
    \includegraphics[width=0.45\textwidth]{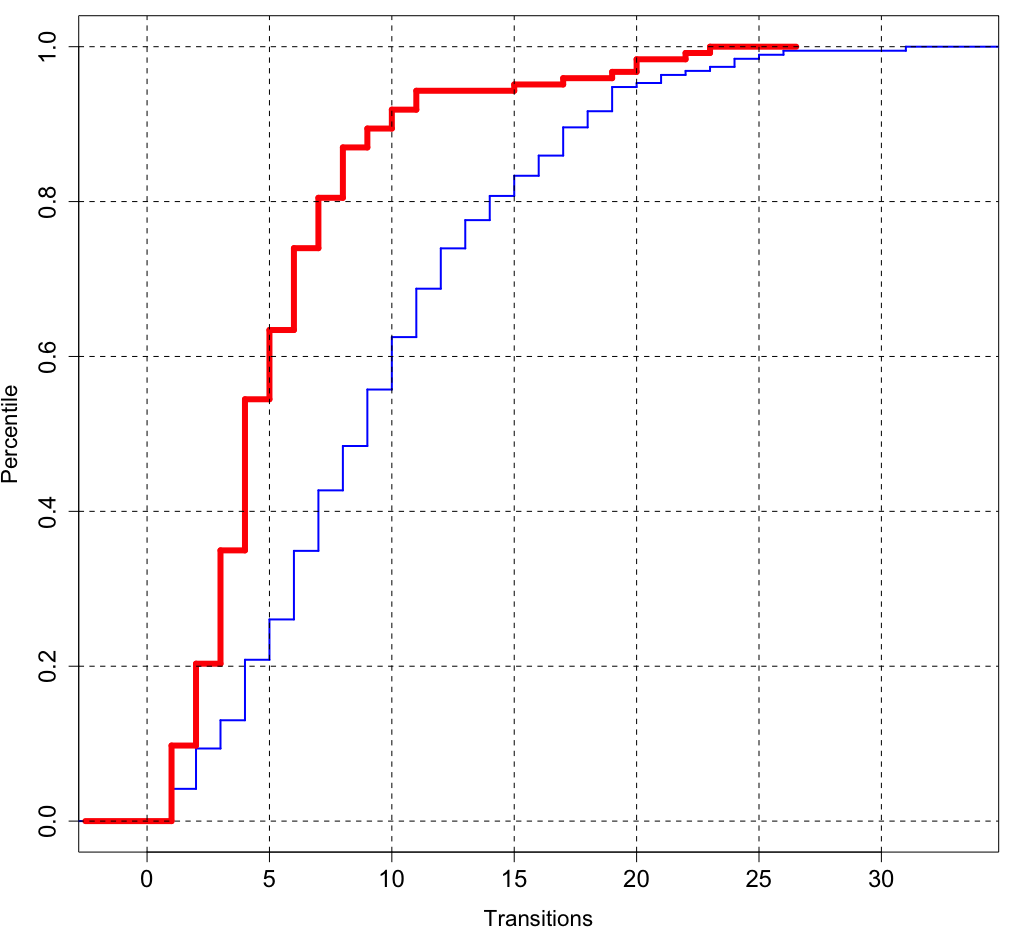}
    \caption{CDF of unique transition locations made by occupants of \adorms, plotted from pre \virus time ($P_U0$ the blue line on the right) to the current quarantine phase ($P_U1$ the red line on the left). At least 50\% of the occupants had reduced their transitions from approximately 10 unique locations to 5.}
    \label{fig:umassDormCDF}
\end{figure}

We next investigated if similar changes to the mobility rate had occurred due to the quarantine policy.  Figure~\ref{fig:umassDormCDF} plots the CDF of the number of other places visited by the occupants of one particular \amherst dorm (called \adorms) in each phase. We observed that the number of visits had reduced with the $50^{th}$ percentile reducing by slightly more than half (about ten visits in $P_U0$ versus less than five visits in $P_U1$) and the $90^{th}$ percentile decreasing from about 17 visits to about nine visits. This reduction in mobility behavior is consistent with \cite{badr2020association}, which found that the change in mobility patterns using cell mobile data to reduce by approximately 50\% with state-level policy enforcement.

\begin{figure}[h]
    \centering
    \includegraphics[width=0.6\textwidth]{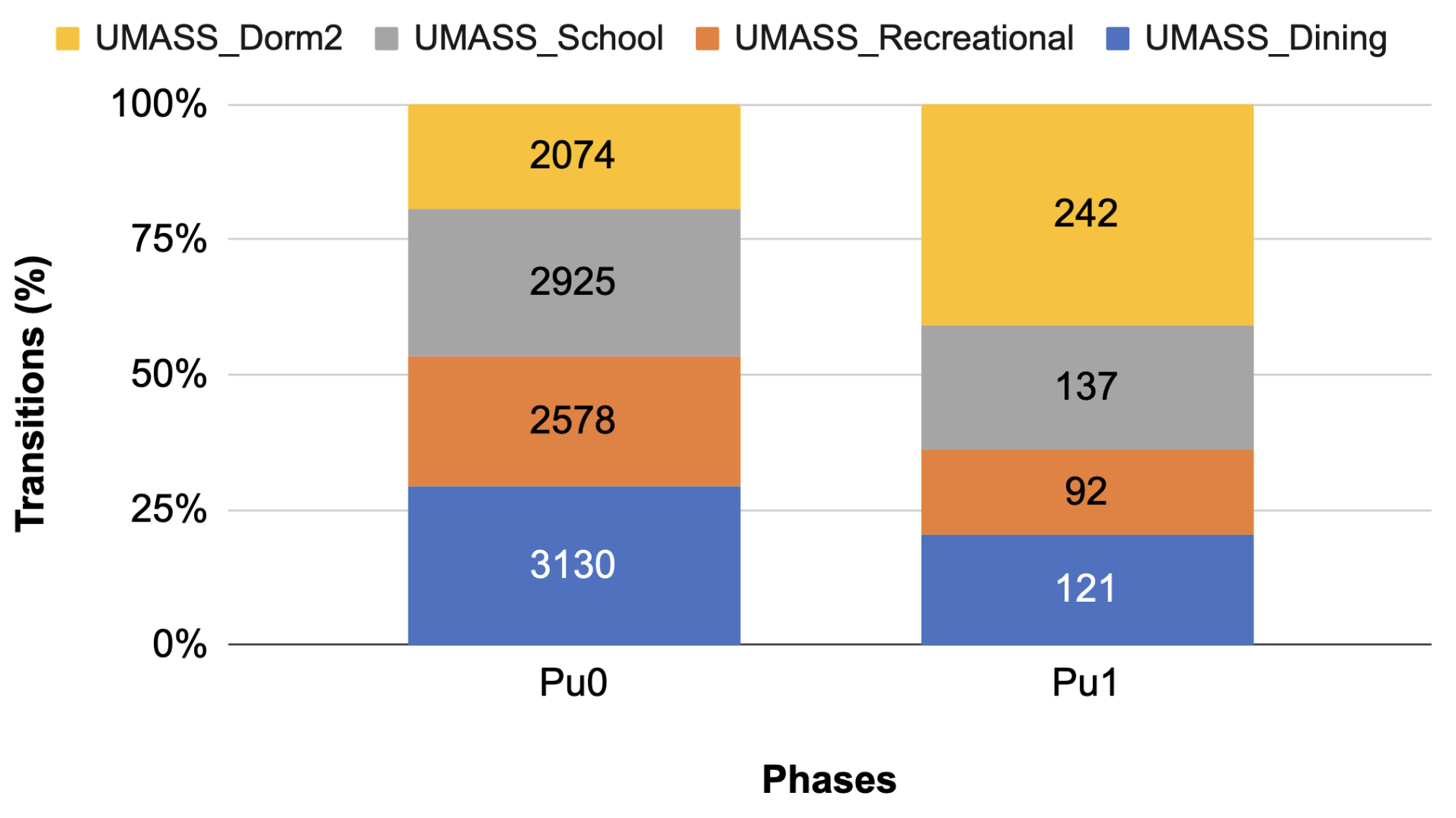}
    \caption{The percentage change in transitions for different areas made by occupants of \adorms, plotted from pre \virus time ($P_U0$) to current time ($P_U1$).}
    \label{fig:umassDormOthers}
\end{figure}

Figure~\ref{fig:umassDormOthers} breaks down these visits by the type of place visited. We observed that most of the visits in $P_U0$ were to dining and recreational locations. However, in $P_U1$, most of the visits were made to another on-campus dorm. We believe this was attributed to the students availing themselves of the dining and recreational facilities as certain previously-popular places on campus had been closed in $P_U1$.

\section{Policy Impact}
\label{sec:discussion}
This section discusses the impact of the policy decisions on disease spreading amongst the campus community. In particular, we looked at two different modes by which the virus could spread: 1) spreading amongst the people in the same place as where they are located. We call this vector {\em local spread}. 2) spreading amongst intercrossing people as they travel to and spend time at places other than the primary location. We call this vector {\em mobility spread}. Controlling each of these vectors requires different approaches. 

Controlling {\em local spread}
requires reducing the density of people in the same location. On the other hand, controlling {\em mobility spread} requires reducing the amount of movement outside one's primary location. Reducing the density of people, in general, can reduce {\em mobility spread} as well. Note: both these approaches can apply in both cases except with different intensities. For example, limiting the movement one does in one's primary location can control {\em local spread} but this may not always be practical.
  
\subsection{Controlling {\em Local Spread} Across Campus}

From the results presented in Sections \ref{sec:analysis_sin} and \ref{sec:analysis_us}, we note that quarantine policies were very effective in removing people from their workplaces. The policies immediately removed one primary source of {\em local spread} (i.e., spreading a virus amongst co-workers). 

However, this quarantine policy resulted in higher densities being observed in the student dormitories, as shown in Figure \ref{fig:umassOverall} where the occupancy in dorms increased after the initial quarantine measures were imposed.  This was ``solved'' by the universities asking students to vacate the dorms and return home. This reduced the density of the dormitories as shown by our results, but it moved the problem elsewhere. In particular, we believe that many home residences would have seen much higher population densities as a result of these quarantine policies as the entire country (Singapore) and state (in the US) population was asked to stay at home for extended periods of time. This could make it easier for residents to fall sick if someone in their vicinity had the virus. 
Indeed, Singapore experienced this first-hand as the second wave of \virus outbreak in Singapore occurred in the dormitories used by foreign workers. The population densities at these dormitories were very high, and the first cases of \virus were reported on April 1st 2020~\cite{Workers_Dorm_First}. This spread grew very fast and resulted in thousands of infections within the dormitories within just a few weeks~\cite{Workers_Dorm_Large}. Fortunately, the strict quarantine policies, enacted just a few days (on 7th April, 2020) after the first dorm infection (on 1st April, 2020) when the authorities realized the potential for the spread to grow out of control, ensured that the virus was contained within the dormitories. For example, while the infection rate remains high in the worker dormitories, the number of cases in the rest of Singapore is almost non-existent -- on May 15th, 2020, 791 new infection cases were discovered in the worker dorms with just one other case discovered across the rest of Singapore~\cite{SingaporeCurrent}. Thus, while quarantine policies can lead to higher local transmission rates, the significantly reduced mobility stops the virus from spreading beyond the local area.

\subsection{Controlling {\em Mobility Spread}}

The second mode of virus propagation is where an infected individual travels to another place and infects someone there. We call this {\em mobility spread}. This vector is particularly dangerous as it can allow the virus to spread to formally safe areas very quickly. Indeed, it was this vector that was responsible for spreading \virus throughout the world -- carried by infected individuals travelling between countries.

We see from the results in Section~\ref{sec:analysis_sin} that only a strict quarantine was effective in reducing mobility patterns. In particular, the split team and other approaches used in phases $P_S1$ and $P_S2$ did not have a significant impact on the mobility patterns of individuals (defined as the number of unique places visited by an individual in a day). However, when strict quarantine policies were enacted, starting in $P_S3$ and fully enacted in $P_S4$, we note that the amount of individual mobility has significantly reduced. Also, when people were mobile, they spent significantly lower amounts of time at each place visited.

This data backs up the policy decisions in both Singapore and the US state to enact a strict quarantine as the impact on individual mobility is very clear. Such measures, in turn, dramatically reduce the probability that \virus can spread beyond a local area. However, as stated previously, reducing mobility comes at the expense of increasing the population density of homes, dormitories, and other residential areas. Hence, there could be a higher probability of local infections as a result of a strict quarantine (as demonstrated by the worker dormitories outbreak in Singapore).

\section{Discussion}
Our study's objective was to demonstrate that coarse-grained WiFi data can sufficiently reflect the spectrum in crowd change when different \virus control policies were implemented. Here we discuss
the implications of our findings.

\subsection{Support Operational Strategies for Universities}
As institutions seek to re-open in the coming semesters, they must fully implement the required \virus protocol and maintain campus operations with as much safety as possible for staff and students. At a higher level, such operational planning encompasses strategies for regulating gatherings in enclosed spaces and tracking infection spread, achievable through monitoring crowd changes using WiFi sensing. Specifically, as shown in Figures \ref{fig:buildOcc} and \ref{fig:US_mobility}, reports on actual space utilization can emphasize over-utilized areas and buildings’ floor-by-floor foot traffic. Monitoring mobility as per Figure \ref{fig:mobilityBuilding} can emphasize possible transmission routes within and among buildings in the campus vicinity. As a whole, monitoring occupancy and mobility can aid institutions backtrack path and areas to focus on disinfecting compromised sites. Other strategies relate to regulating restrictions for external parties and planning for emergency evacuation. The ability to distinguish user types on the WiFi network, as shown in Figure \ref{fig:assocDeassocEvents}, can inform institutions of overcapacity visitors. Monitoring total building capacity remains relevant in reducing the risk of occupant exposure to infection spread in shared open-spaces. For example, establishing one-way safe travels within buildings in the event of an emergency evacuation.

\subsection{Practical Application for Institutional Crowd Controls}
We have since developed an open-sourced monitoring platform that allows institutions to set occupancy parameters based on their own risk assessment for infection spread and mobilize contact tracing within our university campuses: \textbf{\href{https://github.com/umassos/elastic-wifitrace}{https://github.com/umassos/elastic-wifitrace}}

\begin{figure}[t!]
    \centering
    \includegraphics[width=0.8\textwidth]{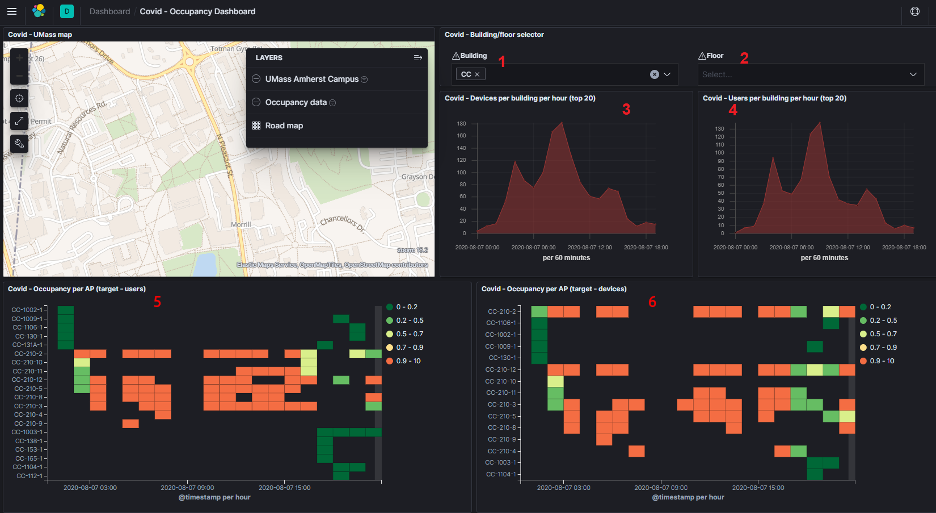}
    \caption{A dashboard visualization of occupancy rate in areas per floor of a building.}
    \label{fig:dashboard}
\end{figure}

Figure \ref{fig:dashboard} is a dashboard visualization of users at \amherst university, driven from their WiFi network device activity. University administrators can select a building and its respective floor to narrow down monitoring to a specific area (e.g., an open study area); see Filter 1 and 2. As explained in Section 3.3, we used the number of unique MAC addresses seen by our WiFi network (devices) to calculate occupancy percentages (Pane 3 and 4). The ability to refine our monitoring parameters and results from per building to areas per floor can quickly help administrators determine areas that may violate zone restrictions. For example, as per Pane 5 and 6, the occupancy heatmaps for each area within the building show a relatively high occupancy rate (in orange) for five areas throughout the day. Note: each area is represented as a row in the y-axis, while the x-axis denotes the time of day.

From an operational perspective, institutions can make more informed decisions to actively revise the restricted capacity based on the severity of an outbreak or official safety compliance guidelines. Alerts on over-utilized spaces help administrators appropriately deploy officers on the ground to manage the crowd. Attention to under-utilized buildings can also be an opportunity for scheduling events in separate physical locations while complying with building occupancy limitations.

\begin{figure}[h!]
    \centering
    \includegraphics[width=0.76\textwidth]{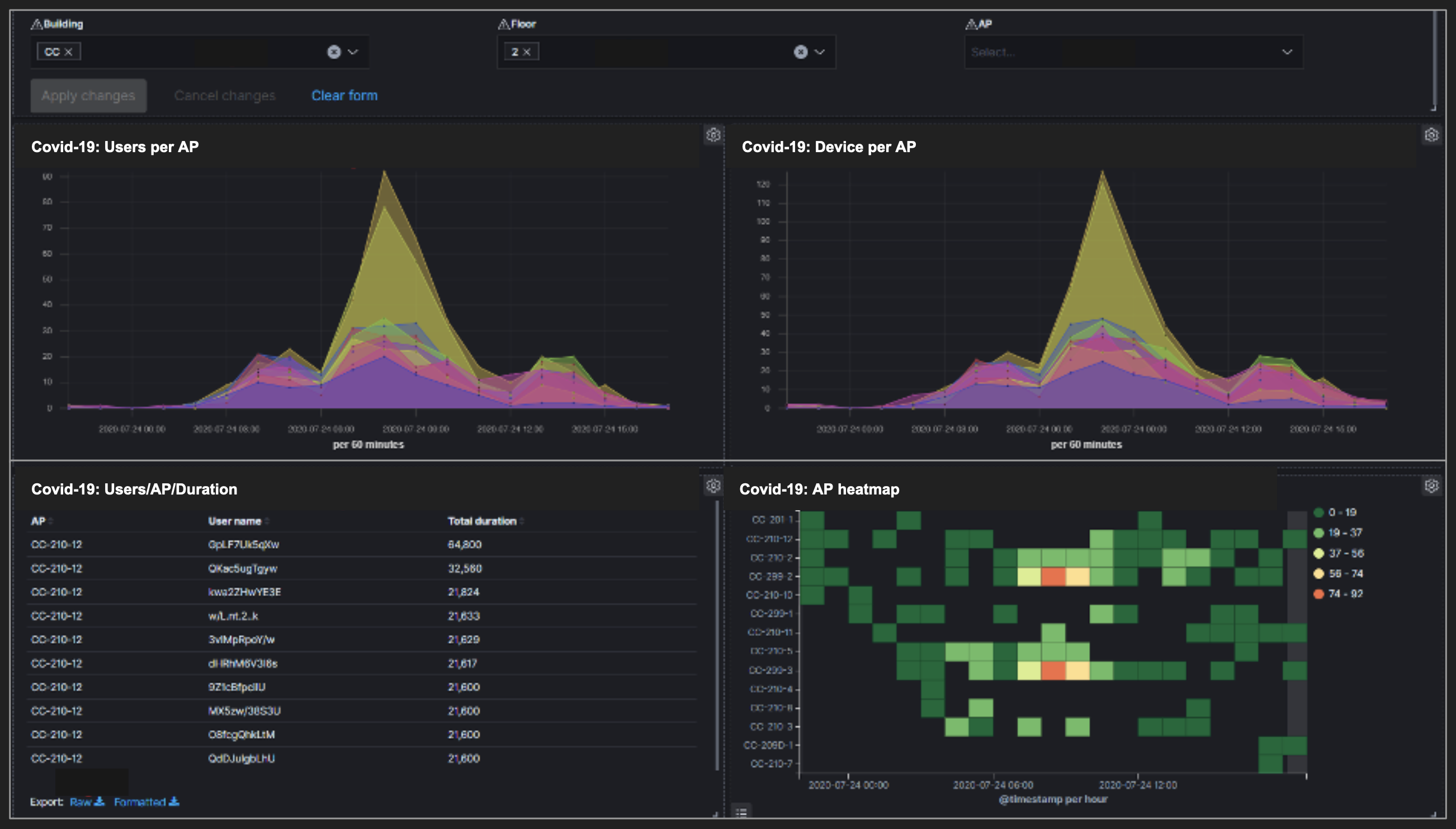}
    \caption{User dashboard displaying all relevant information of occupancy density based on a target location. Left bottom pane lists anonymized occupants.}
    \label{fig:dashboard2}
\end{figure}

\subsection{Implementation of Privacy Safeguards} 
Our goal is towards establishing a non-privacy invasive modality to monitor indoor traffic and implement disease control policies. Recall in  Section \ref{sec:privacyEthicalConsiderations}, we described our privacy safeguards in the event an area is determined to be at high risk of \virus spread, and contract tracing may need to be performed. As shown in Figure \ref{fig:dashboard2} (bottom left pane), we have implemented Elastic WiFiTrace to utilize anonymized WiFi data for measuring occupancy. The mechanism of de-anonymizing this information is recommended for university administrators to determine occupants at-risk of potential exposure. We suggest a standard mandatory notice and consent provision is provided to users of the campus WiFi network to allow for WiFi-based contact tracing. Upon consent, a user preemptively authorizes administrators to access key information, particularly their MAC ID and username, to initiate manual contact tracing procedure.\newline

\noindent Overall, we believe passive WiFi sensing is a promising technique to pursue \virus response on a larger scale. We are already providing updates at \smu to the campus facilities managers and the deans of students at \nus regarding the occupancy and mobility levels across the buildings and dorms at each campus.
\section{Conclusion \& Future Work}
\label{sec:conclusion}
Amid this pandemic, we can anticipate increasingly vulnerable situations to arise as individuals congregate in groups. This paper presented results from two campuses in Singapore and one in the North-Eastern portion of the United States of America, demonstrating how WiFi network information alone could reveal occupancy and mobility spectrum changing with different containment policies. Indeed, augmenting WiFi data can assist institutions with managing safety compliance as we get through the crisis. Our open-source \virus monitoring tool is available: \textbf{\href{https://github.com/umassos/elastic-wifitrace}{https://github.com/umassos/elastic-wifitrace}}. 

This study is not without limitations. First, the data was only collected from WiFi-enabled devices that were associated with the campus WiFi networks; this can result in under-counting if individuals do not enable or carry a WiFi device or are not connected to the campus network. Second, WiFi sensing does not produce fine-grained spatial measures, including occupancy in small rooms that share WiFi APs and inter-user distances. Thus, the modality cannot directly determine individuals who stand too close to an infected person and regard them as at-risk of exposure. Instead, it identifies occupants' degree of risk from being in an exposed area (e.g., an hour-long meeting in a conference room or a lecture classroom), which remains relevant to institutions conducting follow-up manual contact tracing. Third and finally, our data comes from two countries with contrasting \virus policies. While we have observed similar results, local factors may prevent specific results from applying to other regions. As \virus continues to spread, our efforts progress towards deeper and broader analyses addressing these shortfalls.

\begin{acks}
This research is supported by NSF grants 1763834, 1836752, 2021693, 2020888, 2105494 and the Singapore Management University Lee Kong Chian Fellowship.
\end{acks}

\bibliographystyle{main_covid}
\bibliography{main_covid}

\end{document}